\documentclass[12pt]{iopart}

\expandafter\let\csname equation*\endcsname\relax 
\expandafter\let\csname endequation*\endcsname\relax
\usepackage{amsmath} 
\usepackage{amssymb} 
\usepackage{color}
\usepackage{graphicx}
\usepackage{cancel}
\usepackage{enumitem}
\usepackage{cite}
\usepackage{footmisc} 
\graphicspath{{images/}}

\newcommand{\bx}{\mathbf{x}}
\newcommand{\bR}{\mathbf{R}}
\newcommand{\brho}{{\mbox{\boldmath$\rho$}}}

\definecolor{knq}{RGB}{0,0,0}
\definecolor{mca}{RGB}{0,0,0}
\definecolor{mkt}{RGB}{0,0,0}
\definecolor{bbm}{RGB}{0,0,0}
\definecolor{jps}{RGB}{0,0,0}

\newcommand{\mca}[1]{\textcolor{mca}{#1}}
\newcommand{\mkt}[1]{\textcolor{mkt}{#1}}
\newcommand{\bbm}[1]{\textcolor{bbm}{#1}}
\newcommand{\jps}[1]{\textcolor{jps}{#1}}

\begin{document}

\title[Information geometry for multiparameter models]{Information geometry 
\jps{of} multiparameter models: New perspectives on the origin of simplicity}


\author{Katherine N.\ Quinn}
\address{Center for the Physics of Biological Function, Princeton University, Princeton NJ}
\ead{knquinn@princeton.edu}

\author{Michael C.\ Abbott}
\address{Department of Physics, Yale University, New Haven CT}
\ead{michael.abbott@yale.edu}

\author{Mark K.\ Transtrum}
\address{Department of Physics and Astronomy, Brigham Young University, Provo UT }
\ead{mktranstrum@byu.edu}

\author{Benjamin B.\ Machta}
\address{Department of Physics and Systems Biology Institute, Yale University, New Haven CT}
\ead{benjamin.machta@yale.edu}

\author{James P.\ Sethna}
\address{Department of Physics, Cornell University, Ithaca NY}
\ead{sethna@lassp.cornell.edu}


\begin{abstract}
Complex models in physics, biology, economics, and engineering are often
\jps{{\em sloppy}, meaning that the model parameters are not well determined
by the model predictions for collective behavior. Many
parameter combinations can vary over decades without significant changes in the predictions.} 
This review uses information geometry to explore sloppiness and its deep relation to emergent theories. We introduce the {\em model manifold} of predictions, whose coordinates are the model parameters. Its {\em hyperribbon} structure explains why only a few parameter combinations matter for the behavior. We review recent rigorous results that connect the hierarchy of hyperribbon widths to approximation theory, and to the smoothness of model predictions under changes of the control variables. We discuss recent geodesic methods to find simpler models on nearby boundaries of the model manifold -- emergent theories with fewer parameters that explain the behavior equally well. We discuss a Bayesian prior which optimizes the mutual information between model parameters and experimental data, naturally favoring points on the emergent boundary theories and thus simpler models. We introduce a `projected maximum likelihood' prior that efficiently approximates this optimal prior, and contrast both to the poor behavior of the traditional Jeffreys prior. We discuss the way the renormalization group coarse-graining in statistical mechanics introduces a flow of the model manifold, and connect stiff and sloppy directions along the model manifold with relevant and irrelevant eigendirections of the renormalization group. Finally, we discuss recently developed `intensive' embedding methods, allowing one to visualize the predictions of arbitrary probabilistic models as low-dimensional projections of an isometric embedding, and illustrate our method by generating the model manifold of the Ising model.

\end{abstract}

\maketitle

\tableofcontents{}

\markboth{}{} 

\bigskip 
\noindent 

\section{Introduction}
\label{sec:Introduction}

\mkt{
Multi-parameter models in the sciences often contain many more parameters than can be accurately estimated from available data.
In spite of the enormous uncertainty associated with such models, scientists regularly use them to draw physical inferences and make accurate predictions.
This review summarizes recent progress toward resolving this apparent contradiction.
The key observation is that in many scenarios, only a few important parameter combinations are constrained by data and are useful for making predictions,
\jps{and the predictions of the model are correspondingly constrained to an 
effectively low-dimensional hypersurface.
The fact that most complex models exhibit a hierarchy of parameter importance is known as ``sloppiness''\cite{Brown:2003ew,Brown:2004kt}, and the 
hierarchical model manifold~\cite{amari2000methods,Amari2016,nielsen2020elementary,nielsen2022many}
 of predictions is deemed a ``hyperribbon''.}
The emergent simplicity of the model predictions, as represented by the
effectively low dimensional hyperribbon, has broad implications in a variety
of fields.
A previous review addressed to biologists~\cite{Daniels2008} considered the implications of sloppiness for evolvability and robustness of complex biological systems~\cite{draghi2010mutational, tian2011origins}, and more recent work by others has examined similar connections in neuroscience~\cite{o2013correlations, o2015computational, ori2018cellular}.
Writing to chemists~\cite{Transtrum:2015hm}, we emphasized the challenges of estimating tens to thousands of parameters from experimental data and the advantages of simple, effective theories for modeling complex systems.
More broadly, there is growing need for efficient parameter reduction methods that relate detailed mechanistic models to emergent, simpler effective theories~\cite{Transtrum:2014hr, constantine2015active, maiwald2016drivinmodel, froehlich2018efficient, holiday2019manifold, frankle2019lottery, raman2017delinparam}, and others have written about the connections between sloppiness, parameter identifiability, experimental design, uncertainty quantification, and other approaches to large-scale parameter estimation~\cite{brouwer2018underlying, chis2016relationship, dufresne2018geometry, apgar2010sloppmodel, lill2019local,lamont2019correspondence}.
Here, we address the physics community and focus more on the geometrical study of multiparameter models and the perspective it gives on the role of parsimony in mathematical modeling.}

\mkt{
Conventional statistical wisdom is that adding more parameters to a model is only beneficial up to a point.
Eventually, additional parameters are expected to lead to worse predictions as the inferred values become sensitive to the noise rather than signal.
This intuition is formalized by many results from theoretical statistics, such as the Bayes~\cite{BayesInfo} or Akaike~\cite{Akaike} information criteria that explicitly penalize models proportionally to the number of parameters because of their capacity to over-fit.
These results are formally derived in the limit of abundant, informative data, known as the asymptotic limit in theoretical statistics.
But this does not match the reality of scientific inquiry in which experimental data is always limited, and in which science always contains many unobserved details.
For example, systems biology effectively applies models with tens to thousands of rate and binding constants, and in machine learning, the ability of neural networks to generalize from training sets to test sets keeps growing even with millions of parameters \cite{kaplan2020scaling}.
The information geometry methods we review here provide a tangible, direct, and explicit explanation of the predictive power of models with many parameters far from asymptotic assumptions.
The hierarchical nature of parameter importance in sloppy models reflects a low effective-dimensionality.
This means that new parameters are typically unable to drastically improve model fits, and can therefore be added without harming model predictions.
Conversely, most parameter combinations can vary so wildly without affecting
predictions, making them infeasible to measure.
A reduced model description may still be preferred as being more comprehensible, reliable, or likely, but not because the fuller model will overfit.}

In information geometry we view the space of all possible model predictions as forming a manifold, whose co-ordinates are the model's parameters.
There is a natural notion of distance on this manifold, arising from the distinguishability of the predictions from different parameter choices.
And it is an empirical fact that, for a great many models from various branches of science, this manifold has a characteristic shape we term a {\em hyperribbon}.
This means its size in different dimensions varies widely, over many orders of magnitude, and the spectrum of these lengths is approximately evenly spaced on a log scale.
The most important eigendirections need not align with our parameter directions, but their existence is the reason that simpler models can work well.
Section~\ref{sec:ModelManifolds} reviews these ideas, and Fig.~\ref{fig:sloppyEigs} presents evidence that hyperribbons are ubiquitous.



Next we turn to the question of why this hyperribbon structure appears in so many systems, including those already known to admit much simpler ``spherical cow'' effective models.
In Section~\ref{sec:HyperribbonBounds} we show rigorously, for a widely used class of models (nonlinear least-squares models with certain smoothness approximations), that the range of possible model predictions indeed is bounded in such a way that forces our hierarchical hyperribbon structure, implying that models with tens to thousands of control knobs will nonetheless have behaviors that could be captured by a much simpler theory~\cite{Quinn:2018tw}.


In many cases different simpler theories may appear in different limits.
The most famous examples of this are the two reductions of 20th century physics to 19th century physics, by either turning off quantum effects, or turning off relativity.
These classical theories can be viewed as living on edges of the full manifold, each with one less constant of nature, that is, one less dimension.
We discuss such edges in Section~\ref{sec:EffectiveTheories}; one can generate these simpler theories for nearly any model by exploring the hierarchy of edges of the model manifold. 
One can remove parameter combinations by taking limits at which their specific values become unimportant, thereby obtaining a simpler model. 
We provide an algorithm~\cite{Transtrum:2014hr} for systematically generating these reduced models using geodesics on the model manifold. 
For models from diverse applications ranging from power networks to systems biology~\cite{Transtrum:2016jm,transtrum2017measurement,Niksic:2016qau,bohner2017identifiability,lombardo2017systematic}, these methods have not only succeeded in generating dramatically reduced models, but the resulting models vividly illustrate key features of the complex underlying networks and match the same simplified models that experienced scientists have devised.


For Bayesian analysis of uncertainty in high-dimensional models, the volume of the manifold starts to matter.
Just as high-dimensional spheres have most of their surface area near an equatorial line, so most of the volume of the model manifold is far from the edges where the simplified models dominate.
The most common statistical criteria for model selection are justified in
the limit of no uncertainties (excellent data). This leads to a weight (the {\em prior})
that is uniform over the volume in prediction space -- wildly distorting model selection in multiparameter cases by actually selecting against the simpler boundary models.
In Section~\ref{sec:Priors} we discuss two new priors inspired by our information geometry analysis, both of which place extra weight on the manifold edges using only the anticipated uncertainty of future experimental data.
On manifolds with many thin dimensions, our new priors lead to much better performance, but depend very much on the amount of data to be gathered.
The first prior is obtained by maximizing mutual information~\cite{Mattingly:2017uao, Abbott-2020-WIP}, an old idea whose consequences with finite information were long overlooked.
While this is optimal, it is a discrete distribution, and not easy to solve for.
The second prior (not published elsewhere) we call projected maximum likelihood, and is designed to retain many of the good properties while being much easier to sample from.



There are other principled approaches to simplifying theories, the most famous of which in physics are continuum limits and the renormalization group.
These slowly discard complex microscopic behavior, to arrive at quantitatively accurate models for the long length-scale behavior -- whether in condensed
matter physics or quantum field theory.
In Section~\ref{sec:ParameterCompression}, we discuss the relationship of these methods to information geometry. For example, coarse-graining
causes the manifold to be compressed along irrelevant dimensions, while the relevant dimensions are unchanged.
This can be seen both locally, on the eigenvalues of the Fisher metric ~\cite{Machta:2013ga}, and globally by studying the induced flow on the model manifold~\cite{Raju:2017ty}.


Finally, the low-dimensional emergent geometries predicted by information geometry lend themselves to direct visualization of model manifolds. 
For least-squares fits to data, visualization of model manifolds can be done using principal component analysis (PCA), which determines a projection into two or three dimensions based on the fraction of variation explained. 
Because of the hierarchical nature of the hyperribbon structure of model manifolds, PCA can effectively display the emergent low-dimensional model behavior
in least-squares models.
We use this method to illustrate hyperribbon bounds (Fig.~\ref{fig:Parallelotope}) and the advantages of our new maximum likelihood and projection priors (Figs.~\ref{fig:Comparison-Jeffreys-26D} and~\ref{fig:Mark-Jeffreys-26D}).
In Section~\ref{sec:MinkowskiSpace}, we extend this visualization method
using replica theory and Minkowski-like embedding spaces to address more general models that predict non-Gaussian statistical distributions. There we provide two visualizations of interest to physicists:
 the two-dimensional Ising model from statistical mechanics and the 
cosmological $\Lambda$CDM model for the fluctuations in the cosmic
microwave background radiation.


\section{The information geometry of sloppy models}
\label{sec:ModelManifolds}

Information geometry is where differential geometry meets statistics and information science.
Models and data have geometric structures which reveal underlying relationships.
For instance, the models discussed in Section~\ref{sec:EffectiveTheories} form high-dimensional manifolds that can be embedded in a space of possible model predictions. 
Their connections to one another are revealed in the way their manifolds relate to each other.
Furthermore, the phenomena of ill-posed and well-constrained parameter combinations (sloppy and stiff) manifest geometrically as short and long directions along the model manifold.


The simplest scenario for a geometric interpretation are \jps{{\em nonlinear
least-squares}} models which, given input parameters, produce a prediction or output as a high-dimensional vector. Often these involve time: The model is some nonlinear function $y_\theta(t)$ with $D$ model parameters $\theta = \{ \theta^\mu \}$ evaluated at points $(t_1,t_2,\ldots)$. We call the set of all possible model predictions for all possible model parameters the model manifold $\mathcal{Y}$:
\begin{align}
\label{eq:ManifoldDef}
\mathcal{Y} = \left\{Y({\theta})\vert{\theta}\right\}\quad \textrm{where} \quad Y({\theta}) = [y_\theta(t_1),y_\theta(t_2),\dots, y_\theta(t_M)].
\end{align}
The intrinsic dimension of the manifold is (at most) the number of input parameters, $D$. We describe it as embedded in a space whose dimension, $M$, is determined by the number of predictions made. If the model predicts one number $y_\theta(t_i)$ at each time $t_i$, then this construction embeds the manifold in Euclidean space, $\mathcal{Y} \subset \mathbb{R}^M$. \jps{(The general definition for
the model manifold is given in Section~\ref{subsec:LocalAndGlobal}.)}

\begin{figure}
\center
\includegraphics[width=\textwidth]{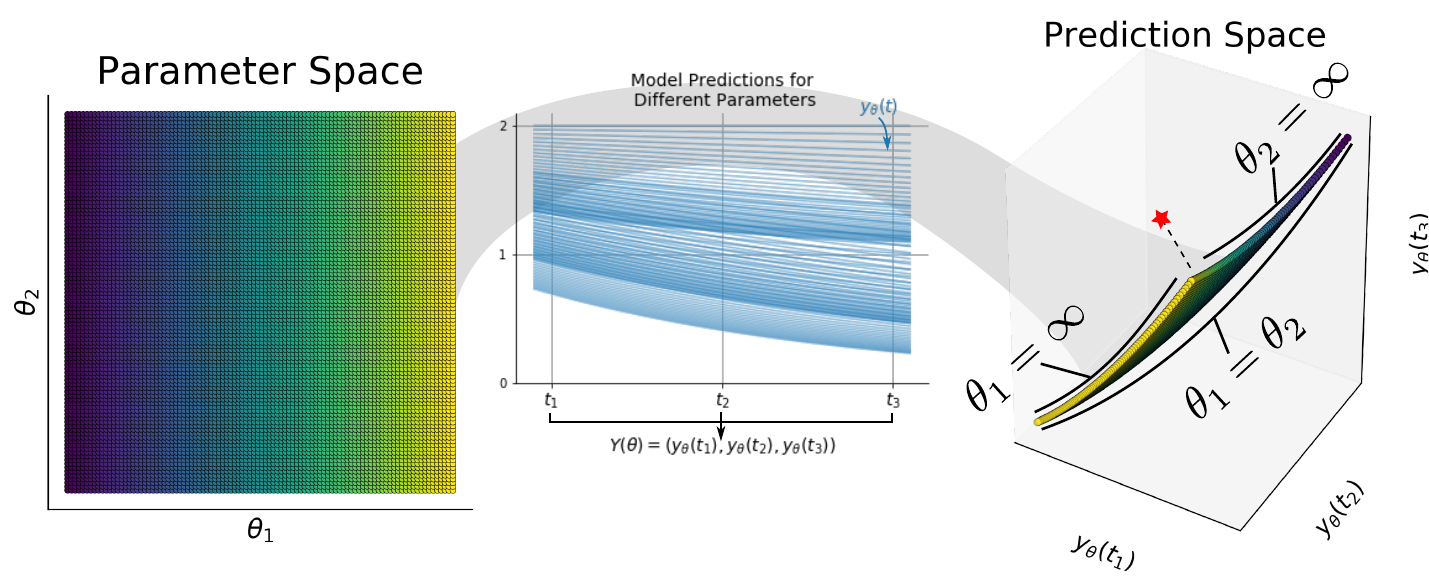}
\caption{Model manifolds for a toy nonlinear model, the sum of two exponential curves with different decay rates $\theta_1, \theta_2 \geq 0$ (from Eq.~\ref{eq:trivialModel}). Model manifolds are constructed as the set of all possible predictions for all possible model parameters and seen as a mapping from parameter space to prediction space. Fitting model parameters to data can be interpreted as projecting data (the red star) onto the manifold.}
\label{fig:cartoonManifold}
\end{figure}

As an illustration of this, consider a classic nonlinear model of decaying exponentials with two input parameters $\theta_1$ and $\theta_2$:
\begin{align}
\label{eq:trivialModel}
y_\theta(t) = e^{-\theta_1 t} + e^{-\theta_2 t},\quad \theta_\mu \geq 0.
\end{align}
Suppose that we evaluate this at three points in time, $(t_1, t_2, t_3)$. Then as we sweep through the different parameter values, we generate a curved 2D manifold embedded in a flat 3D space, as shown in Fig.~\ref{fig:cartoonManifold}. The geometry of this manifold reflects properties of the model itself: There is a symmetry to the model, since $\theta_1$ and $\theta_2$ can be interchanged without changing the prediction, and these points are identified in the manifold, since it folds on itself about \smash{$\theta_1 = \theta_2$}. The boundaries represent important limits of the model, such as when $\theta_1$ or $\theta_2$ go to infinity (and thus the corresponding exponential goes to zero), or when \smash{$\theta_1 = \theta_2$}. Note that boundaries are lines, 1D sub-manifolds, corresponding to one fewer degree of freedom. These represent the simpler models obtained by taking the corresponding limit.%
\footnote{
\mkt{Our ``model manifold'' is not strictly a manifold. The interior of 
the model manifold fits the definition: every neighbourhood has $D$ co-ordinates in the usual way. But the important simpler models
at the boundary introduce fold lines, corners, and other singularites that
violate the formal mathematical definition of a manifold with boundaries.}
}

Fitting model parameters to data has an elegant interpretation in this framework: It can be seen as projecting data onto the model manifold, as illustrated in Fig.~\ref{fig:cartoonManifold}. The red star is the measured data \smash{$(x_1, x_2, \dots, x_M)$}, viewed as a point in the embedding space of the model manifold. To find the parameters at which the vector of predictions $[y_\theta(t_1), y_\theta(t_2), \dots, y_\theta(t_M)]$ is closest to the data point, we should minimize a sum of squares,%
  \footnote{\jps{Hence the name `nonlinear least-squares' model.}}
\begin{align}
\label{eq:LSchi2}
\chi^2 =\sum_{i=1}^M \frac{\left(y_{\theta}(t_i) - x_i\right)^2}{2\sigma_i^2}.
\end{align}
Minimizing this measure is identical to choosing the maximum likelihood parameter, for a probabilistic model in which the observed data point $x$ is the prediction $y$ corrupted with Gaussian noise. This may be written $x = y + \mathcal{N}(0, \sigma^2)$, or more explicitly
\begin{align}
\label{eq:gauss_x_y}
p(x \vert \theta) = \frac{1}{(2\pi\sigma^2)^{M/2}} \exp\left(\frac{-\sum_i(x_i-y_\theta(t_i))^2}{2\sigma^2}\right)
\end{align}
Then $\chi^2 = -\log p(x \vert \theta)$ is precisely the cost being minimized, (Eq.~\ref{eq:LSchi2}), up to an irrelevant additive constant.

In this interpretation, the fitting of model parameters to data always involves assumptions about the noise.
None of our measurements are perfectly precise, and ignoring this just means building in unstated assumptions. 
Instead, we interpret every model as a probabilistic model for what data we will see, $x$.
Given an outcome $y$ as a deterministic function of the inputs, $\theta$ and $t_i$, we explicitly incorporate the experimental noise to generate a predicted
probability distribution \jps{$p(x|\theta)$} around $y$.
This allows us to extend what we have said about least-squares model manifolds to apply to more general models.

We now turn to how we measure distance on the model manifold.
The embedding $\mathcal{Y} \subset \mathbb{R}^M$ described above is special
because it is {\em isometric}: the Euclidean $\mathbb{R}^M$ distances between
points in $\mathcal{Y}$ locally agree with the
natural measure of statistical distance, the Fisher metric.

\subsection{Local and global distances}
\label{subsec:LocalAndGlobal}

To judge whether different parameter values are similar, or points on the model manifold are nearby, we need some notion of distance.
The idea of information
geometry~\cite{amari2000methods,Amari2016,nielsen2020elementary,nielsen2022many}
is that this must come from the model itself, and the noise in our measurements: If two different values predict data which is easily distinguishable, then we should not regard them as being near to each other.
Conversely, if two different parameter values are close together, they must predict nearly indistinguishable data.

\jps{For nonlinear least-squares models, the distance between the predictions
of two parameters $\theta$ and $\tilde\theta$ is given by the $\chi^2$
distance of Eq.~\ref{eq:LSchi2},
$\sum_{i=1}^M (y_{\theta}(t_i) - y_{\tilde\theta}(t_i))^2/2\sigma_i^2$, which
is basically the Euclidean distance between the two vectors $y_\theta$ and
$y_{\tilde\theta}$. For more general models, distinguishability is measured between the probability distributions $p(x\vert\theta)$ predicted for a set of 
possible outcomes $x$ 
(spin configurations for the Ising model~\cite{Teoh:2020kp},
temperature maps for the cosmic background radiation~\cite{Quinn:2019dc},\dots).
We shall discuss three such measures, valid for general probability
distributions, in Section~\ref{sec:MinkowskiSpace}: the 
Hellinger distance~\cite{hellinger1909neue} (Eq.~\ref{eq:Hel-defn}),
the Bhattacharyya divergence~\cite{bhattacharyya1946measure}
(Eq.~\ref{eq:Bhat-defn}),
and the Jeffreys~\cite[p.158]{jeffreys1939theory} or symmetrized Kullback--Leibler~\cite[Eq.~2.5]{kullback1951information} divergence
(Eq.~\ref{eq:symkl}).}

\jps{All of these measures are zero when $\theta = \tilde\theta$, are symmetric,
and can be calculated for any two points $\theta$ and $\tilde\theta$; they are 
thus a global measure of distance. All of these measures reduce to the $\chi^2$
Euclidean distance for the vector predictions of least-squares models.
The latter two measures do not obey the triangle
inequality, but each corresponds to distances in a Minkowski-like
space where some directions are `time-like' (Section~\ref{sec:MinkowskiSpace}).
These three, together with a wide family of other divergences~\cite{csiszar2004information,renyi1961measures}, share the same local measure,} 
that is, the distance over an infinitesimal change $\tilde\theta = \theta + d\theta$:
\begin{align}
\label{eq:FIM}
ds^2 = \sum_{\mu,\nu=1}^D g_{\mu\nu} d\theta^\mu d\theta^\nu, 
\quad
g_{\mu\nu}(\theta) = \sum_{x} \frac{\partial \log p(x \vert \theta)}{\partial \theta^\mu}  \frac{\partial \log p(x \vert \theta)}{\partial \theta^\nu} p{({x} \vert {\theta})}.
\end{align}
This is the Fisher metric, or $g_{\mu\nu}$ for $\mu,\nu=1,2,\ldots,D$ the \emph{Fisher information matrix} (FIM) \cite{Fisher:1922jk}.
It may be thought of as measuring distance in parameter space in units of standard deviations, using the width of the distribution $p(x\vert\theta)$ on the space of possible data. \jps{(Indeed, the famous Cram{\'e}r-Rao bound~\cite{rao1945information,Cramer,Rao} uses the Fisher information metric to bound the ability to distinguish the
predictions between two nearby models -- the ability to know that 
the parameters $\tilde \theta$ are incorrect given data generated by parameters
$\theta$.)  We shall study these more general model manifolds for 
probabilistic models by generating {\em isometric embeddings}. By preserving
one of the distance measures
$d^2(\theta,\tilde\theta)$ between every pair of models $\theta$ and $\tilde\theta$, we generate a 
geometrical representation of the model manifold that both preserves the FIM and
(in the cases studied so far) exhibits the same sloppy hyperribbon structure as 
we have found in least-squares models.}


\begin{figure}
\center
\includegraphics[width=0.6\textwidth]{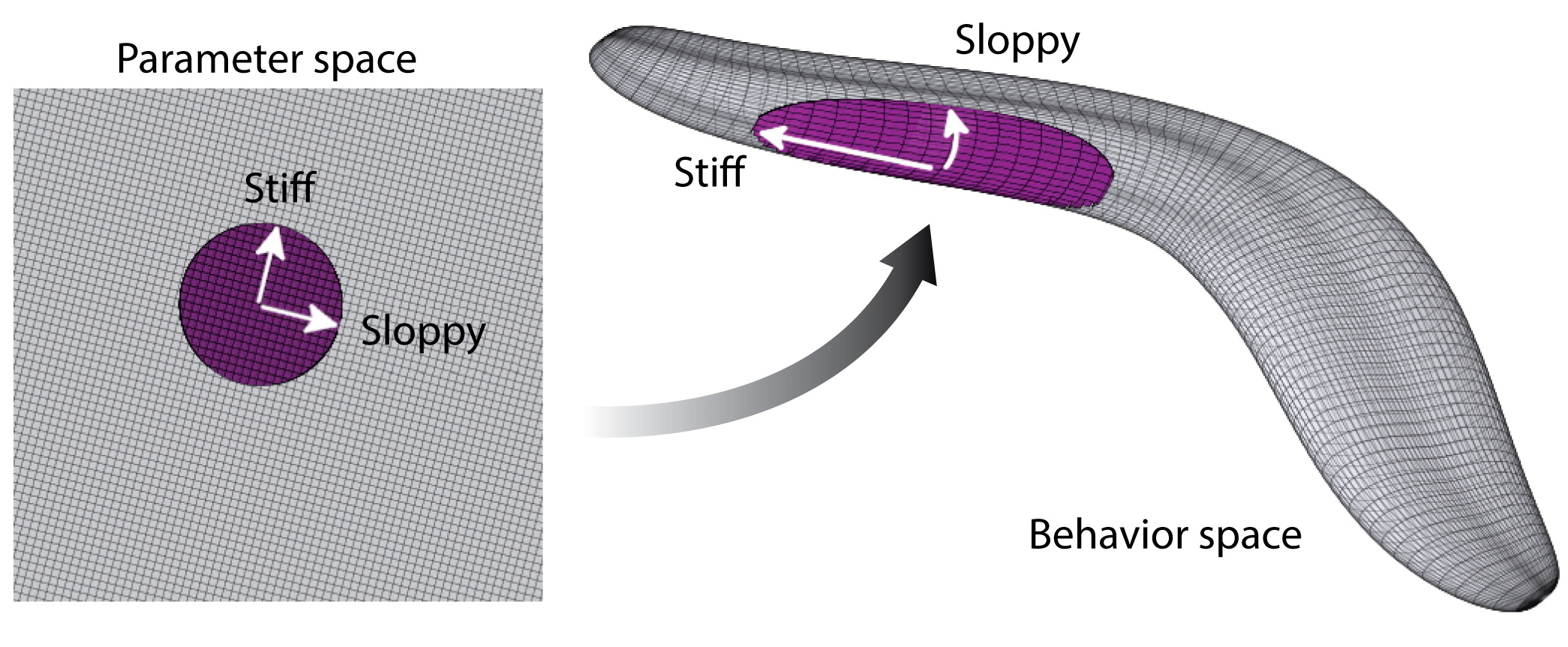}
\caption{The prediction/behavior space of complex systems is primarily controlled by a few parameter combinations, even when the model has numerous parameters.
The hierarchy of parameter importance directly translates to a geometric hierarchy in the local properties (the eigenvalues of the metric roughly form a geometric series) and global features (the manifold widths have a similar hierarchical decay). Figure taken from~\cite{Sethna2017}.
}
\label{fig:crystalSloppy}
\end{figure}

While individual components of the FIM tell us how sensitive the model is to particular parameters, we are more interested in its sensitivity to parameter combinations.
In the extreme case, many different parameters may control the same mechanism, making the model sensitive to changes in each of them in isolation.
But if they all have equivalent effects, then the model is effectively one-dimensional, and it will be possible to make correlated changes which cancel, and do not affect the results. 
In this case, we would expect the FIM to have one large eigenvalue, and many small or zero ones. 
The one important eigenvector we will call the \emph{stiff} direction in parameter space, and the remaining unimportant ones \emph{sloppy}.
The stiff direction is both important for predictions, and tightly constrained by observed data; the sloppy directions are neither. Fig.~\ref{fig:crystalSloppy} draws this in the context of an embedding $\mathcal{Y} \subset \mathbb{R}^3$.

\begin{figure}
\center
\includegraphics[width=\textwidth]{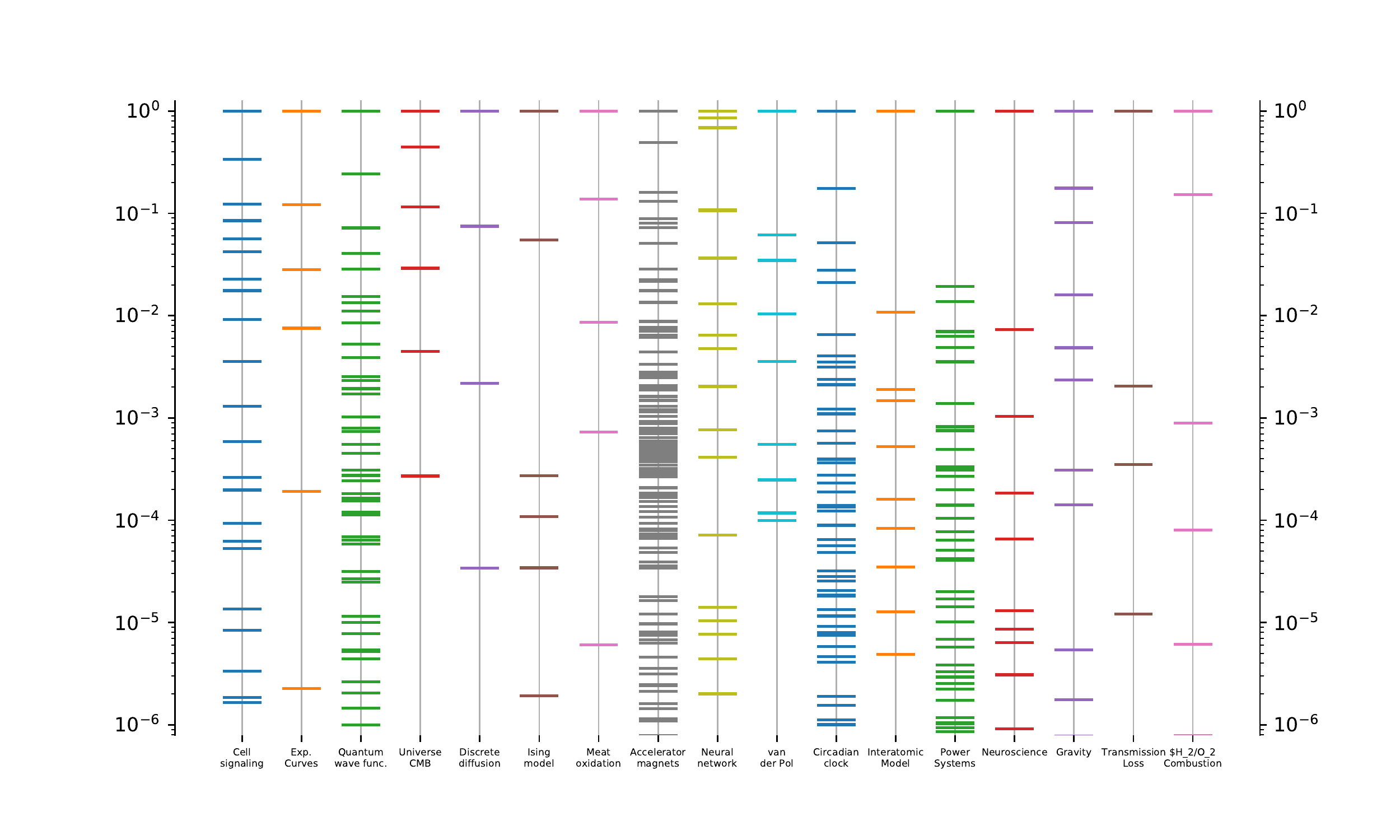}
\caption{Parameter hierarchies that span many orders of magnitude are observed in a wide variety of models.
The vertical axis shows eigenvalues of the Fisher information matrix, scaled by the largest eigenvalue.
Cell signaling data from~\cite{Brown:2004kt}, radioactive decay and neural network are taken from~\cite{Transtrum2011}, quantum wavefunction are taken from~\cite{Waterfall2006}, diffusion model and Ising model are taken from~\cite{Machta:2013ga}, meat oxidation is from~\cite{Tofteskov2019}, CMB data from~\cite{Quinn:2019dc}, accelerator model taken from~\cite{Ryan2007}, van der Pol oscillator taken from~\cite{Chachra2012}, circadian clock model from~\cite{Daniels2008}, SW interatomic potential model from~\cite{wen2017force}, power systems from~\cite{transtrum2018simultaneous}, gravitational model from~\cite{daniels2015automated}, transmission loss in an underwater environment from~\cite{mortenson2021parameter}, and finally $H_2IO_2$ combustion model from~\cite{PepiotPC}.
}
\label{fig:sloppyEigs}
\end{figure}

In more realistic models with many parameters, the eigenvalues of the FIM often have a characteristic spectrum: They are spread over many orders of magnitude, and are roughly evenly spaced on a log scale. 
Fig.~\ref{fig:sloppyEigs} shows some examples of this, from explicit computation of the FIM for models from a wide variety of sciences. 
It is a striking empirical fact that so many of them display this structure. 
\jps{These eigenvalues tell us about the local geometry of the model manifold, that is, the effect of small changes in the parameters.
While these are extremely suggestive, they have the flaw that they depend on the particular parameterization chosen.
If we re-scale some parameters, such as by changing their units, then the FIM and hence the spectrum of eigenvalues will change.}%
\footnote{\jps{In fact, the process of diagonalizing the FIM to find its eigenvectors amounts to finding a basis for parameter space, near a point, in which the matrix becomes $1$.}
}

\jps{We can avoid this by measuring the global geometry of the model
manifold~\cite{Transtrum:2010ci,TranstrumMS11} using {\em geodesics} along the 
thick and thin directions. Let $\theta(\tau)$ be a path in parameter space,
which then is mapped by the model onto a path on the model manifold. The
length of this path, as defined by the Fisher information metric
\begin{equation}
L = \int \sqrt{ds^2} = \int_0^1 d\tau \sqrt{\smash{\sum_{\mu\nu}} g_{\mu\nu}(\theta(\tau)) \frac{\partial\theta^\mu}{\partial\tau} \frac{\partial\theta^\nu}{\partial\tau}}
\end{equation}
is independent of the way we define the coordinate system $L$ on the model
manifold. A geodesic is a curve connecting two points which minimizes the
length between them.  
For models with Gaussian noise $x = y + \mathcal{N}(0, \sigma^2)$ as in Eq.~\ref{eq:gauss_x_y} above, this geodesic distance agrees with the Euclidean path
length in the prediction space, measured by walking along $\mathcal Y$.}

\begin{figure}
\center
\includegraphics[width=\textwidth]{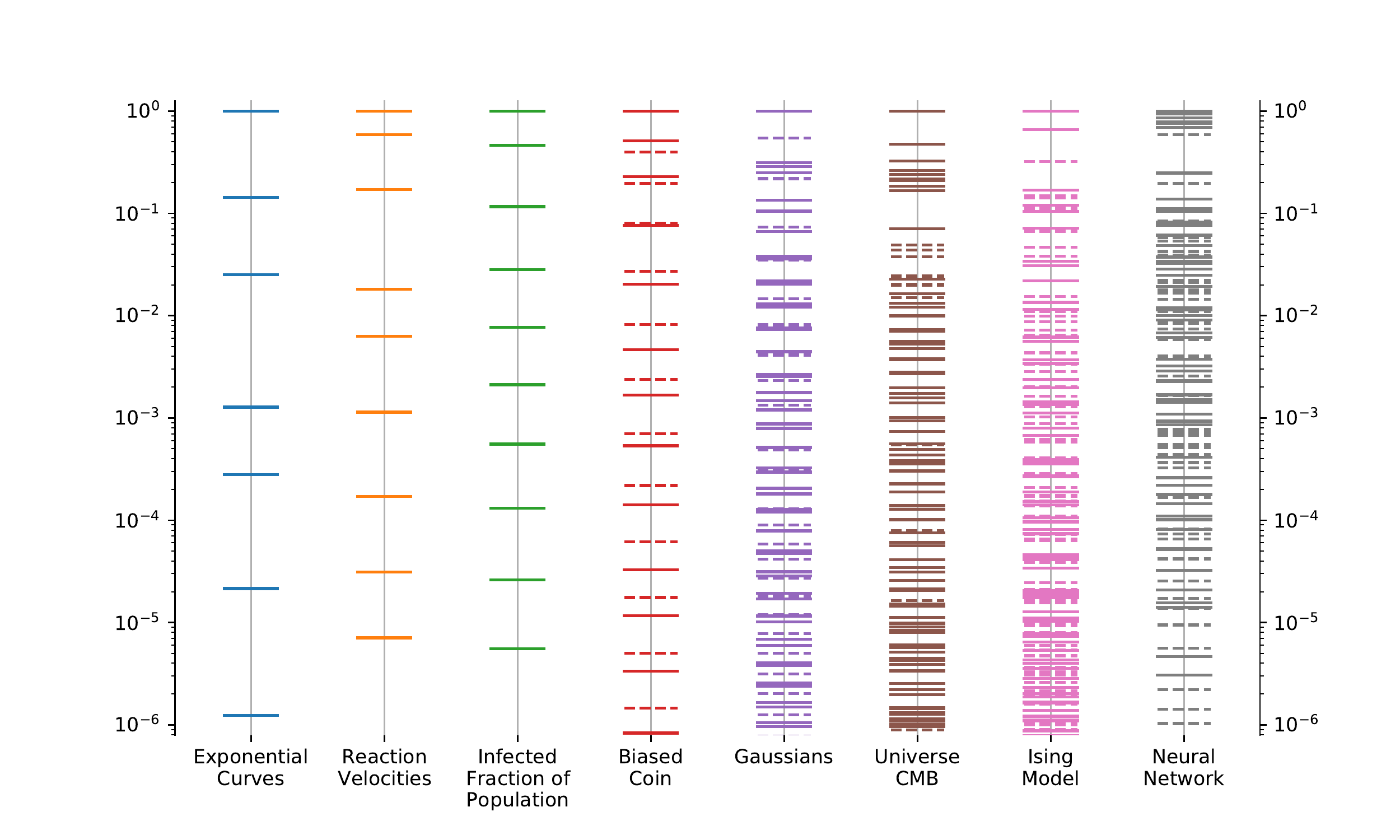}
\caption{Manifold widths for many disparate, nonlinear models (rescaled by the largest width for each model) illustrating the hyperribbon structure that characterizes model manifolds.
Note the enormous range in vertical axis. For probabilistic models, imaginary lengths (\textit{i.e.} negative squared distances) are reflected by dashed lines (Section~\ref{sec:MinkowskiSpace}).
Exponential curves, reaction velocities, and epidemiology model taken from~\cite{Quinn:2018tw}, biased coin, Gaussians, CMB, Ising Model and Neural Network taken from~\cite{Quinn:2019dc}. 
\jps{The first three models here, and models 1-3 and 
7-17 of Fig.~\ref{fig:sloppyEigs}, are nonlinear least-squares models,
hence subject to the rigorous hyperribbon bounds of
Section~\ref{sec:HyperribbonBounds}. The other probabilistic models show
the same hyperribbon hierarchy; methods for visualizing their model
manifolds are discussed in Section~\ref{sec:MinkowskiSpace}. Section~\ref{subsec:isKL} provides a different hyperribbon embedding that yields a
finite-dimensional embedding for all of the latter except CMB (zero widths
beyond the first few).}
}
\label{fig:sloppyLengths}
\end{figure}

\jps{We measure the widths of the model manifold by starting from an initial point
(say, the best fit to the experimental data) and launching geodesics along 
one of the eigendirections in parameter space, shooting both ways until it 
hits the boundaries.  Fig.~\ref{fig:sloppyLengths} shows the resulting geodesic
widths across the model manifold for a number of different models. As advertised, they too show a roughly geometric progression over many orders of magnitude,
with the stiff parameter directions yielding the widest directions and with 
the many sloppy directions corresponding to incredibly thin directions on the
model manifold.
We will call this emergent hierarchical structure of the model manifold a \emph{hyperribbon}, after the idea that a ribbon is much longer than it is wide, and much wider than it is thick.}

\jps{We call a model sloppy, and the model manifold a hyperribbon, if the
parameter-space eigenvalues and prediction-space widths are hierarchical:
roughly equally spaced in log, spanning several decades. A sloppy model 
will have an effective low-dimensional description if many of its widths are
small compared to the precision of the predictions demanded. This need not
be the case -- the $\Lambda$CDM model of the Universe cosmic microwave
background radiation (column~3 in Fig.~\ref{fig:sloppyEigs} and column~6 in
Fig.~\ref{fig:sloppyLengths}) is sloppy (especially considering how few
parameters it has) because the experiments are so precise that all parameters
are well determined from the data.
An impressive use of optimal experimental design~\cite{apgar2010sloppy}
has shown that the individual parameters in our cell signaling network~\cite{Brown:2004kt,Brown:2003ew} (column~1 in Fig.~\ref{fig:sloppyEigs}) could in
principle be determined, albeit with a rather large number of
experiments~\cite{ChachraTS11}. Alternatively, one could measure all the
individual parameters separately in a model whose {\em collective} behavior
is sloppy~\cite{zwolak2005globally}, but the resulting collective behavior
will usually be predictable only if every one of the parameters is well
specified~\cite{Gutenkunst:2007gl}. Finally, most scientific models 
of complex systems like biological signaling are likely approximations to the
microscopic behavior (Michaelis Menten equations replacing enzyme reaction
kinetics), and the resulting parameters are `renormalized' to incorporate
coarse-grained ignored complexity. Systematically designing experiments
to measure these coarse-grained model parameters may just uncover flaws in the
microscopic description that are irrelevant to the emergent behavior. We
shall systematically study these useful coarse-grained models in 
Section~\ref{subsec:MBAM}.}

\section{Hyperribbon bounds \jps{for nonlinear least-squares models}}
\label{sec:HyperribbonBounds}

\newcommand{\nth}{\ensuremath{n^\mathrm{th}}}
\newcommand{\curlyY}{\ensuremath{\mathcal{Y}}}
\newcommand{\curlyF}{\ensuremath{\mathcal{F}}}

A fundamental feature of sloppy models is the hyperribbon structure of
the model manifold. As discussed in the previous section, the model
manifold has a hierarchy of widths that follows a geometric 
spread---roughly even in log. Where does this ubiquitous hierarchy of sensitivity
come from?
In this section we shall review rigorous bounds on the manifold of possible predictions for \jps{the special case of a multiparameter nonlinear least-squares model}~\cite{Quinn:2018tw}. The proof relies on the smoothness of the predictions as the experimental controls (time, experimental
conditions, \dots) are varied, and explains the hierarchy of widths of the
model manifold (as in Fig.~\ref{fig:sloppyLengths}).
The smoothness and analyticity of the functions used in model construction
directly controls the range of possible predictions allowed by the model.
By quantifying the former we can understand the hyperribbon nature of the latter.

Before we start, we should emphasize how the sloppiness we study is distinct
from other well-studied mechanisms that can make systems ill-posed. First, in systems biology and other fields,
many have studied {\em structural identifiability} -- when certain
parameters cannot be gleaned from a given set of experiments, {\em e.g.}
when some perhaps nonlinear transformation of the parameters makes
for exactly the same predictions~\cite{dufresne2018geometry}.
These would give eigenvalues of the
metric exactly equal to zero. Our work instead focuses on what one would call
{\em practical identifiability}~\cite{dufresne2018geometry}, where
some parameter combinations are almost interchangeable, leading to the 
small, sloppy eigenvalues in parameter space.
Second, while the models we
study often have a variety of time or length scales, the observed 
hierarchies are {\em not} primarily due to a separation of
scales~\cite{HOLIDAY2019419}. Systems with a separation of scales
(fast-slow dynamical systems, singular perturbation theory, boundary layer 
theory) have been thoroughly studied in the literature, and certainly
will generate a large range of eigenvalues, but do not explain why
eigenvalues span many decades uniformly. To test that separation 
of scales is not required for sloppiness, one need simply take a classically sloppy model such as a complex network in systems biology~\cite{Brown:2003ew,Brown:2004kt} and set all the rate
constants to one. The cost Hessian
remains sloppy, with eigenvalues roughly uniformly spread over many
decades.
We also note that models with diverging time scales will typically violate
the smoothness assumptions we need to derive our bounds. That is, a
separation of time scales {\em adds to} the hierarchical sloppiness we derive
in this section.

\subsection{Taylor series}

We can understand sloppiness, and give bounds for the
widths of the model manifold that explain the observed hierarchies, through approximation theory. Specifically, we study polynomial approximations of model
predictions. For simplicity, here we consider models whose predictions are 
a one-dimensional function $y_\theta(t)$ over the interval $t \in (-T,T)$,
such as in Fig.~\ref{fig:cartoonManifold}
where we choose $T=0.9$. For models where this is not the case, we can simply rescale $t$ so that it fits in this region. The bounds on model predictions shall rely on the smoothness of $y$ as
the experimental control $t$ is varied. 

In the simplest case, assume $y_\theta$ is analytic in $t$, with Taylor series $\sum a_k t^k$ and coefficients $|a_k| \le 1$ for all $k$. This is a sufficient condition to give the series radius of convergence equal to one, larger than the radius $T$ of the time range being studied. Consider approximating $y(t)$ by truncating its Taylor series. The approximating polynomial, 
$f_{N-1}(t) = a_0 + a_1 t + \dots + a_{N-1} t^{N-1}$, differs from
$y(t)$ by at most
\begin{equation}
\label{eq:epsilonN}
\epsilon_N = 
\sum_{N}^\infty |a_k| t^k  \le
\sum_{N}^\infty T^k 
\le T^N/(1-T),
\end{equation}
since $|a_k|\le 1$ and $|t|<T $. Recall from Eq.~\ref{eq:ManifoldDef} that the model manifold $\mathcal{Y}$ is defined as the space of all possible model predictions $y_\theta$ for varying model parameters $\theta$. Understanding the constraints on $f_{N-1}$ for each $y_\theta$ will therefore allow us to understand what constraints are set on the model manifold $\mathcal{Y}$. Due to the construction of $f_{N-1}$, we have the following linear equation:
\[
\label{eq:VandDef}
\begin{bmatrix}
f_{N-1}(t_0) \\
f_{N-1}(t_1) \\
\dots \\
f_{N-1}(t_{N-1})
\end{bmatrix} 
= 
\begin{bmatrix}
a_0 + a_1 t_0 + \dots + a_{N-1} t_0^{N-1}\\
a_0 + a_1 t_1 + \dots + a_{N-1} t_1^{N-1}\\
\dots \\
a_0 + a_1 t_{N-1} + \dots + a_{N-1} t_{N-1}^{N-1}
\end{bmatrix}
= 
\begin{bmatrix}
1 & t_0 & \dots & t_0^{N-1}\\
1 & t_1 & \dots & t_1^{N-1}\\
\dots & \dots & \dots & \dots\\
1 & t_{N-1} & \dots & t_{N-1}^{N-1}
\end{bmatrix}
\begin{bmatrix}
a_0\\
a_1\\
\dots\\
a_{N-1}
\end{bmatrix}
\]
or, in particular
\begin{equation}
f_{N-1}(\mathbf{t}) = V(\mathbf{t}) \mathbf{a}
\label{eq:polyapprox}
\end{equation}
where $V$ is the well-studied Vandermonde matrix $V_{jk} = (t_j)^k$.
The best-known property of the Vandermonde matrix is its determinant
$\det(V) = \prod_{0\le i \le j < N} (t_i-t_j)$.
Notice that, since $t\in (-T,T)$, there are $\sim N^2/2$ factors
$|t_i - t_j|$ ranging in size from $2 T$ down to $~1/N$: the determinant
(the product of the eigenvalues) will be very small. 

Because the model predictions are characterized by the coefficients in their Taylor series as well as where they are evaluated (\textit{i.e.} the $a_i$ and $V(\mathbf{t})$ in Eq.~\ref{eq:polyapprox} respectively), understanding what constrains both $a_i$ and $V(t)$ will provide an understanding of what bounds the model.

\begin{figure}
\center
\includegraphics[width=0.9\textwidth]{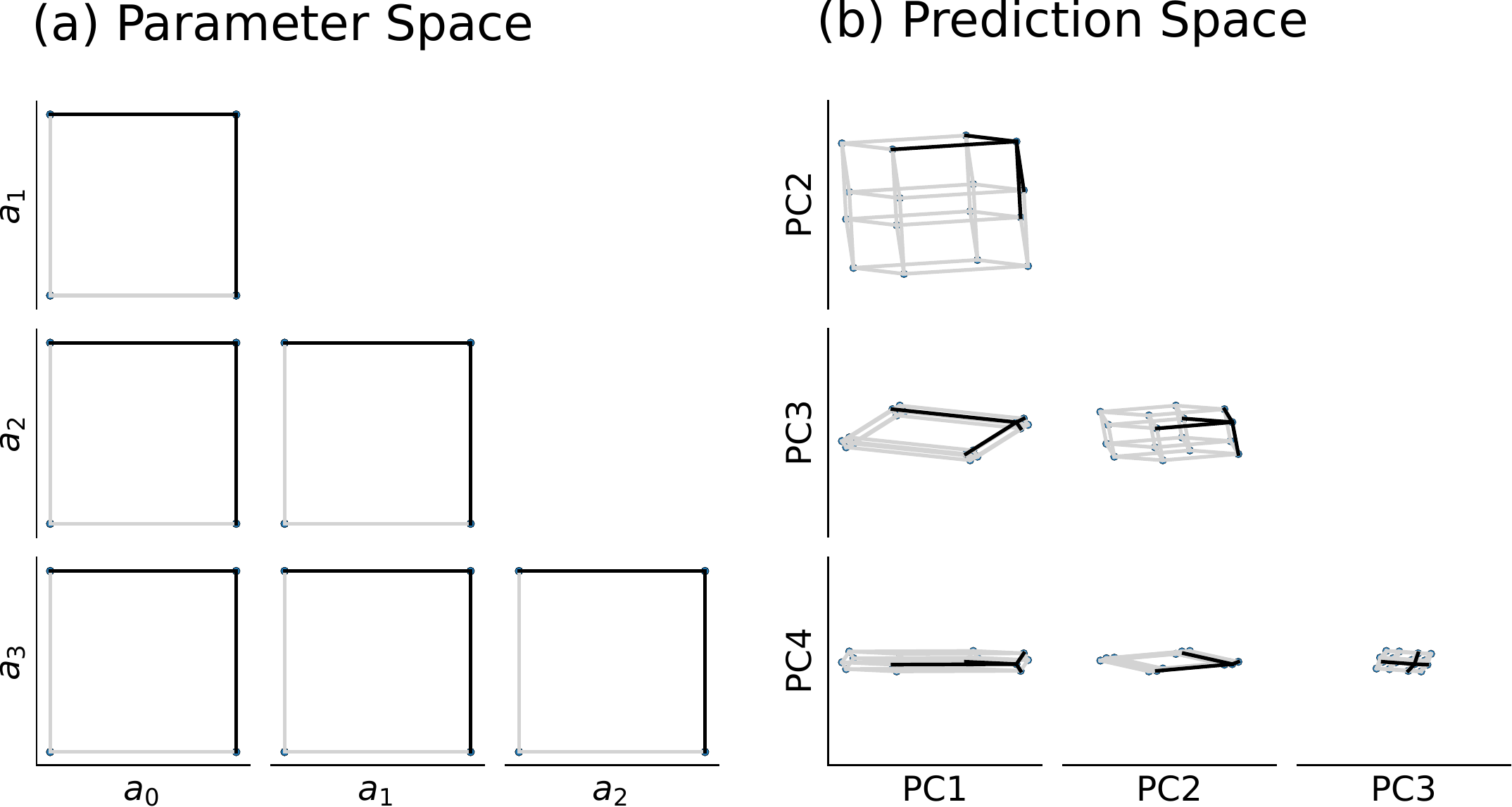}
\caption{\textbf{Model manifold for $\curlyF$, projected along various 
principal axes of $V$.} (a) The coefficients of $f_{N-1}(t)$ are constrained such that $|a_i|\leq 1$. They therefore form a region in parameter space for $\mathcal{F}$ in the shape of a hypercube. (b) When evaluated at points $\mathbf{t} = [-0.9,-0.5,0,0.4,0.8]$, the set of predictions are constrained by a hyper-parallelepiped whose widths are determined by $V$ in Eq.~\ref{eq:VandDef}, which we visualize by projecting along the Principal Components. Because every point in the model manifold $y_\theta$ is at most $\epsilon_N$ away from a corresponding $f_{N-1}$, the shape in (b) constrains the model manifold $\mathcal{Y}$, forcing a flat hyper-ribbon structure. Black lines show how the right angles in a vertex of the hypercube becomes highly skewed in prediction space.
}
\label{fig:Parallelotope}
\end{figure}

For instance, for our models where the $a_i$ are limited to be between -1 and 1, the Vandermonde matrix takes the hypercube of possible values of $\mathbf{a}$ and maps it to a hyper-parallelepiped $\mathcal{F}$ in prediction space that is incredibly skewed, as shown in Fig.~\ref{fig:Parallelotope}; its volume is proportional to the incredibly small determinant.
The model manifold $\mathcal{Y}$ is constrained to be within $\epsilon_N = T^N/(1-T)$ (Eq.~\ref{eq:epsilonN}) of this hyper-parallelepiped because each $y_\theta\in\mathcal{Y}$ has a truncated power series $f_{N-1}$ which differs from it by at most $\epsilon_N$. The highly skewed nature of this hyper-parallelepiped, whose widths are determined by the Vandermonde matrix, forces a similar hierarchical structure on the model manifold $\mathcal{Y}$.

\subsection{Chebyshev polynomials}

\begin{figure}
\center
\includegraphics[width=\textwidth]{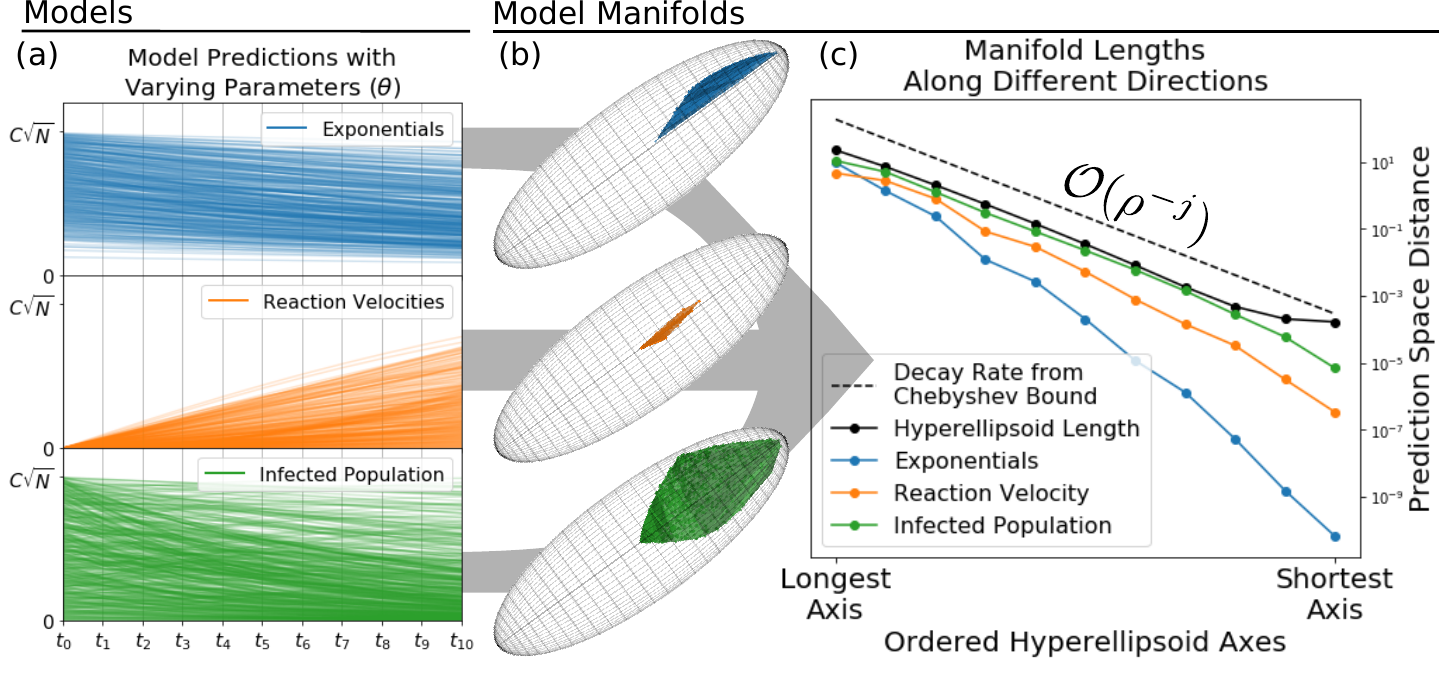}
\caption{\textbf{Model manifold and bounds} of three disparate models: (1) exponential curves such as those in Fig.~\ref{fig:cartoonManifold}, (2) reaction velocities of an enzyme-catalyzed reaction, and (3) the infected population in an SIR model. (a)~Each model is evaluated at specific points $t_i$ for varying parameter values, yielding a set of different predictions. (b)~Each model manifold is bounded by a the same hyperellipsoid, whose widths follow a geometric decay $\mathcal(O)(\rho^{-j})$, calculated from the singular values of the Vandermonde matrix combined with an error from Chebyshev Approximation. (c)~The widths of the hyperellipsoid are plotted against the successive widths of each model manifold, showing a clear hyperribbon structure. The smoothness of the models determines the value of $\rho$, which in turn sets the widths of the hyperellipsoid bound, which forces a hyperribbon structure for the model manifolds. Figure taken from~\cite{Quinn:2018tw}. 
}
\label{fig:ModelBounds1D}
\end{figure}

To expand the understanding of why a hyperribbon structure is observed for non-analytic models, in~\cite{Quinn:2018tw} it is proven for a relaxed smoothness condition where the coefficients $a_i$ are bounded not by a hypercube but by an n-sphere of radius $r$, and not for all coefficients but only those up to a certain order. Bounds were derived both using a Chebyshev approximation and 
by using the Taylor series considered above. Following a very similar line of reasoning, it is possible to show that the model manifold $\mathcal{Y}$ is constrained by a hyper-ellipsoid whose $j^\mathrm{th}$ largest principal axis length is bounded by a power law $\rho^{-j}$. To illustrate this, consider three nonlinear models: (1)~a sum of exponentials, such as in Fig.~\ref{fig:cartoonManifold}, (2)~reaction velocities of an enzyme-catalyzed reaction, and (3)~the infected population in an SIR epidemiology model. In all three cases, the models have varying model parameters which produce different predictions, shown in Fig.~\ref{fig:ModelBounds1D}(a). The model manifolds are all bound by the same hyperellipsoid, shown in Fig.~\ref{fig:ModelBounds1D}(b), and form a hyperribbon structure as the successive widths of the hyperellipsoid follow a geometric decay. 

Hyperribbons have also been observed for models with multiple control variables, for instance those which vary for both time and temperature~\cite[Supplemental Material]{Quinn:2018tw}. Here, a two-dimensional Vandermonde matrix is constructed to understand the bounds in prediction space, producing a similar decay in the widths of the hyperellipsoid that bounds the model manifold with additional degeneracies based on the number of additional control variables.

\subsection{Interpolation theory}

\begin{figure}
\center
\includegraphics[width=0.6\textwidth]{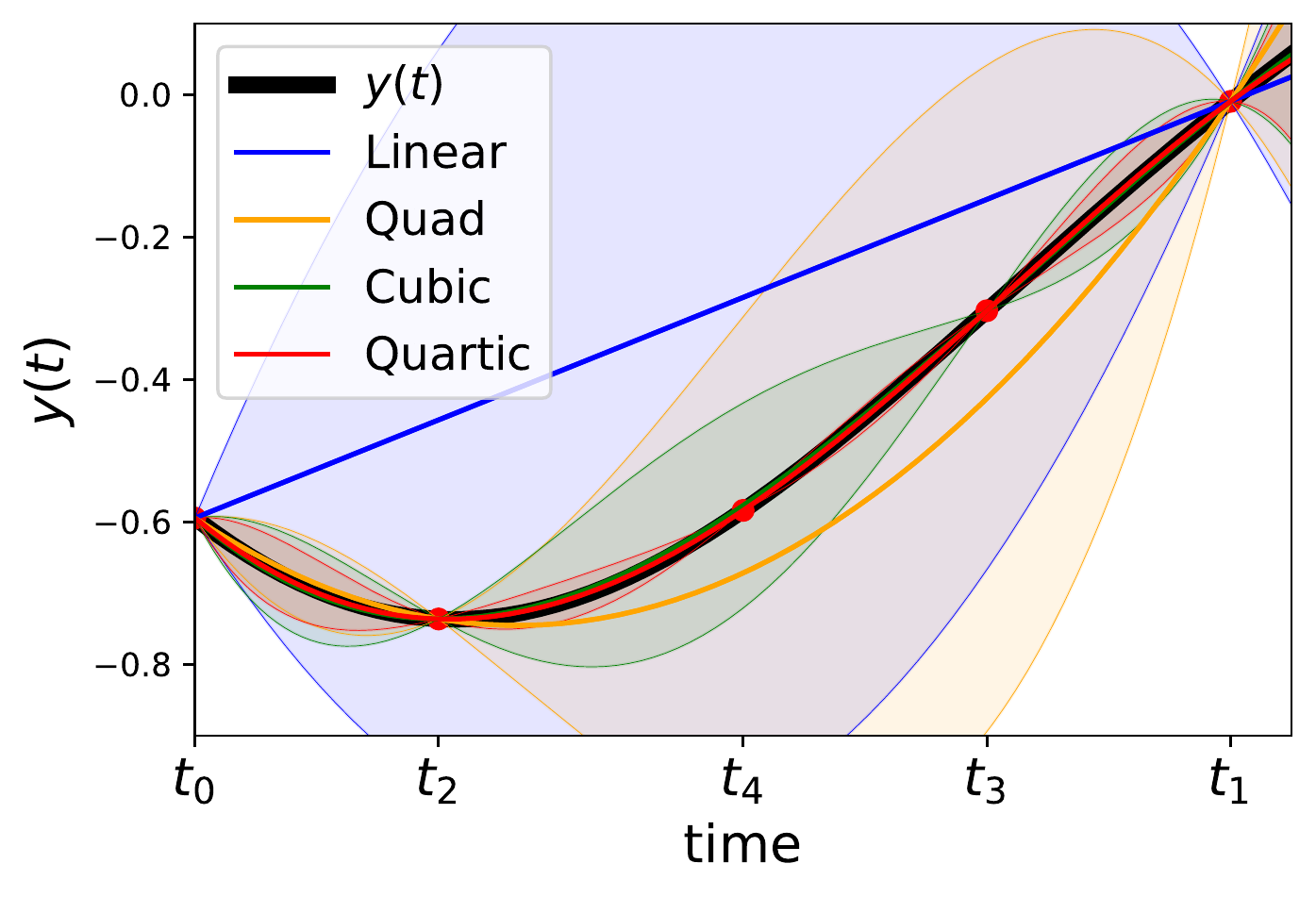}
\caption{
\label{fig:InterpolationFigure}
{\bf Convergence of interpolations.} The black curve is a model prediction for a one-dimensional function $y_\theta(t)$ for $-0.9\le t \le 0.9$. The colored curves are polynomial fits $f_{n-1}(t)$ as we add data points $y_\theta(t_n)$ at new interpolating points $t_n$
Note that the widths of possible errors in the fits (shaded regions)
get smaller as we
add more points: the interpolations converge with more data. The shaded regions give the usual error bound on the interpolation, given the smoothness of $y_\theta(t)$. Our function has $y^{(n)}(t)/n! < 1$ with radius of convergence at any point slightly larger than our interval. The geometrical decrease in the widths of these bounds with $n$ is related to our hyperribbon bounds on the widths of successive cross sections of
the model manifold. 
}
\end{figure}

We can get a different intuitive picture for why smoothness in the control parameters implies a hyperribbon structure in the predictions, by considering interpolation theory~\cite{Transtrum:2010ci,TranstrumMS11}. Consider studying successive cross sections of the model manifold
generated by experimental measurements at times $t_0,t_1,\dots$, here for analytic functions whose radius of convergence $|y_\theta^{(n)}(t)|/n! <1$ for all points $t$ in the interval.~\cite{Transtrum:2010ci,Transtrum2011}. The value of $y_\theta(t)$ at each time $t\in (-T,T)$ is an experimental prediction, and therefore a coordinate axis of our model manifold. Because there are an infinite number of points on the interval, and each point lends itself to a dimension on the model manifold, the resulting manifold is infinite-dimensional. Because of the bounded derivatives,%
  \footnote{The function $y_\theta(t)$ in Fig.~\ref{fig:InterpolationFigure}
   was chosen by generating random quintic polynomials  
   $\sum_{\mu=0}^5 \theta_\mu (t)^\mu$ with bounded monomial coefficients 
   $|\theta_\mu|<1$, and then selecting for ones whose Taylor expansions
   at all other points in the interval have the same bound (hence
   $\max_{t,n} |y^{(n)}(t)|/n!<1$). This further selection allows for
   application of the standard interpolation theorem~\cite{Transtrum:2010ci}. 
   Both conditions correspond to forcing a radius of convergence equal to one, 
   slightly larger than the radius of our interval $t\in(-0.9,0.9)$. But 
   the second criterion, forcing a bounded derivative, leads to a constraint 
   in the allowed space of subsequent predictions which in turn creates 
   the hyperribbon structure.}
measuring $y_\theta(t)$ at a new time further constrains the range of possible values at other times, as shown in Fig.~\ref{fig:InterpolationFigure}.
Each experimental measurement $y_\theta(t_j)$ fixes a point, and has the effect of slicing the model manifold along axis $t_j$ at $y_j$ --- keeping only the parameters $\theta$ whose predictions agree with the experiments.

Figure~\ref{fig:InterpolationFigure} shows a function
$y_\theta(t)\in \curlyY$ taking values $y_\theta(t_j) = y_j$, and rigorous bounds on any function
$z(t)\in \curlyY$ that agree with $y_\theta(t)$ at the constrained time points. The range of predictions $z(t')$ in Fig.~\ref{fig:InterpolationFigure} gives the width of the model manifold cross section after slicing along certain fixed points. Note that this width shrinks with each new experimental measurement, as the effect of additional constraints only further limits the range of allowed functions. Fixing the value of $y_\theta(t_j) = y_j$ shrinks the range of all other possible predictions, just as a ribbon snipped in two has a cut edge that typically reflects the width and thickness of the original ribbon. The convergence of a polynomial interpolation between the points $y(t_j)$ is equivalent to a hyperribbon structure of the model manifold $\curlyY$.

\section{Effective theories as manifold boundaries}
\label{sec:EffectiveTheories}

A remarkable insight that comes out of the geometric study of multi-parameter models is how models of varying complexity relate to one another.
Simple, low-dimensional models often reside on the boundaries of complex, high-dimensional ones.
We have already alluded to this idea for the motivating example in Eq.~\ref{eq:trivialModel}.
To illustrate constructions that will be central to our approach, consider the hierarchy theories to emerge after the discoveries of relativity and quantum mechanics.

To begin, consider a multi-parameter model formulated in the context of special relativity and include the speed of light, $c$, among the parameters.
Varying $c$ defines a sequence of models of co-dimension one.
In the limit of this sequence $c \rightarrow \infty$ the predictions of the model reduce to those of classical Newtonian spacetime.
This is, of course, one of the well-known classical limits and formalizes the relationship between relativity and classical kinematics.
In geometric language, classical kinematics is a boundary of the manifold of special relativities.
Similarly, the limit of $\hbar \rightarrow 0$ identifies classical mechanics as a boundary of quantum mechanics.
Including $G$ as a parameter completes the three-dimensional space of fundamental theories sometimes known as $cGh$ physics~\cite{gorelik2005matvei}.

\begin{figure}
\center
\includegraphics[width=0.55\textwidth]{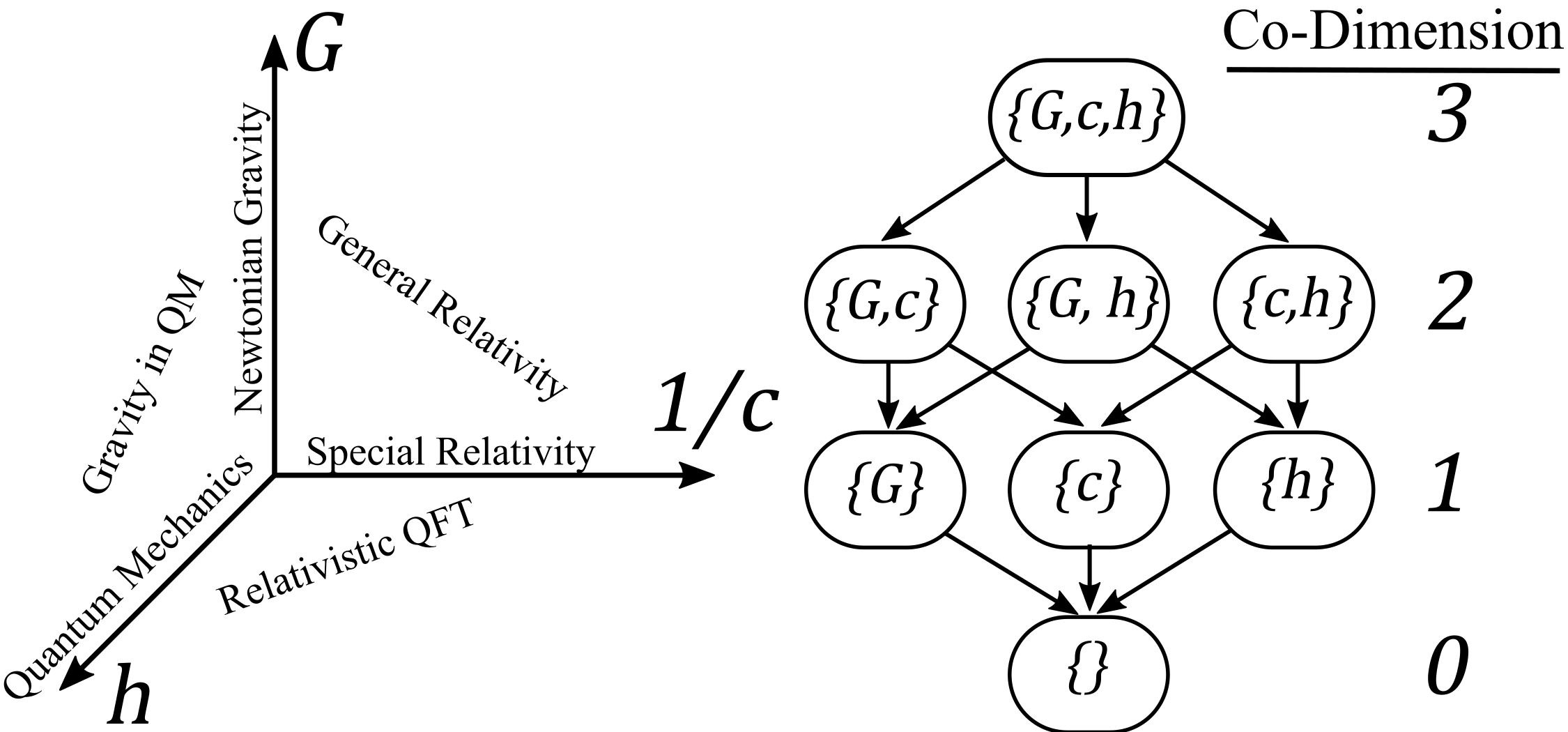}
\caption{The physical constants $G$, $c$, and $h$ parameterize a hierarchy of effective theories connected by a directed graph of limiting approximations.}
\label{fig:newtoneinsteinschrodinger}
\end{figure}

Similarly, manifolds of co-dimension two correspond to models built in the context of general relativity ($\{G, c\}$), relativistic quantum field theory ($\{c, \hbar\}$), and (less well-known) classical gravity in quantum mechanics ($\{G, \hbar\}$)~\cite{kuchavr1980gravitation,padmanabhan2011nonrelativistic}.
These manifolds are bounded by the simpler theories that preceded them and presumably are themselves limiting cases of a more fundamental theory parameterized by all three.
This topological relationship among theories is summarized by the directed graph on the right in Fig.~\ref{fig:newtoneinsteinschrodinger}.

These connection between theories are well-known to physicists and the geometric constructions presented here is not new~\cite{kuchavr1980gravitation,tegmark2008mathematical,padmanabhan2011nonrelativistic}.
Indeed, this relational interpretation of effective theories is a key in formalizing the criteria under which a model is valid and which are likely to be the leading order corrections.
Einstein and Schr\"odinger tell us when and why Newton is right.
Furthermore, this way of organizing scientific knowledge is useful for many types of multi-parameter models.
For example, among continuum models of conventional superconductivity, the detailed theory of Eliashberg involve many greens functions detailing coupling information between electrons and phonons~\cite{eliashberg1960interactions}.
From this, a hierarchy of simplified models emerge as limiting cases, i.e., boundaries of a more complicated model manifold.
The weak-coupling limit gives the theory of Gork'ov~\cite{gor1958energy}, the semi-classical limit gives that of Eilenberger~\cite{eilenberger1968transformation}, and the high-temperature limit gives Ginzburg-Landau theory~\cite{gor1959microscopic}.

Other domains have similar hierarchies of theories and models.
We find this structure generic in multiparameter models.
It motivates an algorithm for systematically constructing effective theories: the Manifold Boundary Approximation Method (MBAM).

\subsection{The manifold boundary approximation method}

\label{subsec:MBAM}

Simplified models are found on the boundaries of the model manifold, and these boundaries exhibit a universal structure.
Returning to the motivating example of two exponentials introduced in Eq.~\ref{eq:trivialModel} and Fig.~\ref{fig:cartoonManifold},
observe that the boundary has a triangle-like structure: A two-dimensional manifold with three one-dimensional edges that meet at three zero-dimensional cusps.
Although the size and shape of the model manifold depend on the observation time, this triangle-like boundary structure is independent of these details and depends only on the form of the observation function, $y_\theta(t)$.
We have noted that each of these boundary elements is a simplified, approximate model.
Formally, this boundary structure is a hierarchical cell complex that is a natural topological decomposition of the boundary.

\begin{figure}
\center
\includegraphics[width=0.35\textwidth]{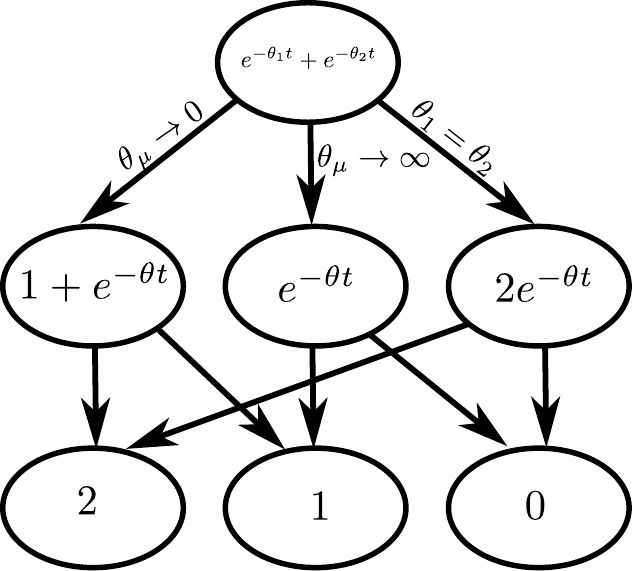}
\caption{The boundary complex of the model manifold is summarized by a directed graph called a Hasse diagram.
  Elements of the boundary complex correspond to simple, reduced models that are related by physically interpretable limiting approximations.}
\label{fig:ExponentialHasse}
\end{figure}

Similar boundary complexes are found in models from very diverse contexts~\cite{Transtrum:2016ahl, Transtrum:2014rma}.
That is to say, a typical $N$-dimensional model is bounded by a collection of $N-1$-dimensional faces, which join at $N-2$-dimensional edges, and so forth.
The adjacency relationship among these facets of different dimensions defines a partially ordered set (often called a POSet), described as graded because each facet has a dimension associated with it.
These relations can be conveniently visualized as a directed graph known as a Hasse diagram, as we have done in Fig.~\ref{fig:newtoneinsteinschrodinger} for $cGh$ theories and repeated in Fig.~\ref{fig:ExponentialHasse} for the exponential example.

Each of the boundary cells of a model manifold corresponds to a simplifying approximation that corresponds to an extreme, idealized behavior of the model.
The Hasse diagram summarizes how these approximations can be composed to yield progressively simpler models.
That is to say, the Hasse diagram of the boundary complex is a road map from a complicated multi-parameter model through various approximations to the distinct behavioral regimes the model enables.
When the model manifold is endowed with an information-theoretic metric, it, along with its boundary structure, is compressed in sloppy directions and stretched in stiff ones.
Useful approximations, i.e., those that retain key explanatory variables, correspond to those boundary cells aligned with the long-axes of the manifold.

We translate this observation into an algorithm, the \emph{Manifold Boundary Approximation Method} (MBAM) that uses information geometry to identify simplified models as boundary cells of a model manifold that is appropriate for a particular data set.
The basic idea is to construct a geodesic shortest-path through the model's parameter space until it intersects a face on the boundary complex. By choosing a sloppy direction in parameter space (searching both directions along an eigenvector of the FIM with small eigenvalue), the geodesic will find a nearby boundary point with nearly identical predictions.
By considering parameter values along the geodesic path, one deduces the limiting approximation that corresponds to the appropriate boundary cell.
Often, this limiting approximation may involve one or more parameters going to extreme values, such as infinite or zero, in a coordinated way.
We re-parameterize the model so that participating parameters are grouped together into a single combination that goes to zero at the boundary of the manifold.
Finally, we analytically apply the approximation to the model to construct a simpler model with one fewer parameter.
We iterate this process until the model is sufficiently simple.

MBAM has a formal similarity to using scale separation to perform model reduction.
Scale separation is one of the oldest and most well-established methods of model reduction.
It introduces a small parameter in which to perturb; the model reduces to a simplified form in the limit that the parameter becomes zero.
Similarly, MBAM determines a combination of parameters, the perturbation parameter, that goes to zero at the boundary of the model manifold.
However, sloppiness adds a new wrinkle to how we use and interpret this procedure.
We do not necessarily require the true value of the perturbation parameter to be small.
Rather, the true parameter value may be large provided the data to be explained do not constrain its value.
Indeed, we observed above that although a model with separated scales will be sloppy, separated-scales are not necessary for sloppiness.
The full model is statistically indistinguishable from the simpler 
separation--of--scales model.
We advocate for using simple models because they reflect the informativity of the data, not because the true system to be modeled has well-separated scales.

\subsection{Relation of MBAM to other reduction methods}

Because the problems related to model complexity transcend scientific disciplines, there are many well-established reduction techniques that have grown-up around specific application areas.
Most of these exploit the specific structure of the problem at hand, usually in conjunction with expert intuition.
For example, dynamical systems are often simplified using singular perturbation theory in which the fast dynamics (often identified by an expert) are replaced by the slow manifold.
Since, MBAM is agnostic to the mathematical form of the model, it can be compared to domain-specific reduction methods.
In many cases, MBAM is formally equivalent to these techniques.
However, because it is a data-driven method, it need not require expert insight and can guide insights into the modeling of complex systems.

For example, consider the problem of modeling transient dynamics of a power system.
Power systems are large, engineered systems with many interconnected components.
Time scales vary from speed-of-light electrical responses to load variations over hours and days.
Models incorporate multi-physics in the form of heterogeneous components that describe not just power delivery, but generation from multiple sources, including renewables, and interactions with economic demand.

Of particular interest are models of synchronous generators, a power systems component that plays an important role in large simulations.
The power community has developed a hierarchy of models describing synchronous generators that are formally connected through a series of singular perturbations.
A typical inference problem involves estimating machine parameters from short disturbances that may last a few seconds.
This problem is challenging because the relevant time scales are intermediate between the fastest and slowest time scales of the system, so it is difficult to judge which time scales are relevant.
Indeed, the true dynamics are not necessarily faster or slower than the observation time scales, but parameters are unidentifiable because of sloppiness.

Applying MBAM to slowly-developing disturbances leads to the expected sequence of singularly perturbed reduced models advocated by theorists.
However, when applied to disturbances at intermediate time scales, the results are more interesting.
Recall that singularly perturbed systems have both an inner and an outer solution corresponding to the fast and slow dynamic respectively.
Rather than uniformly removing fast dynamics, MBAM selects a combination of inner and outer solutions of the singularly perturbed system to remove parameters not constrained by the data.
The remaining parameters are those that could be learned from the data.

Next, we turn to the theory of linear control that has produced many formal insights into the approximation of dynamical systems.
We first discuss how control-theoretic quantities relate to sloppiness, and then demonstrate how MBAM relates to another standard approximation method known as \emph{balanced truncation}.

A linear, time invariant (LTI) system takes the form
\begin{align}
  \label{eq:LTI}
  \dot{x} & = Ax + Bu \\
  y & = Cx + Du \nonumber
\end{align}
where $x(t) \in \mathbb{R}^n$ are state space dynamical variables, $u(t) \in \mathbb{R}^p$ are inputs/control variables, and $y(t) \in \mathbb{R}^m$ are observed variables.
$A$, $B$, $C$, and $D$ are constant matrices of the appropriate shape.
This model arises as the linearization of a generic dynamical system near a stable equilibrium, and theorems proved on LTI systems generalize to nonlinear dynamical systems under certain regularity conditions.
Two important concepts to emerge in the study of LTI systems are observability and controllability.
Roughly, observability is how well the internal state of the system can be inferred from its outputs while controllability is how easily a system can be driven to a particular state through a control protocol.
The concepts of observability and controllability generalize to nonlinear systems, though they are most easily quantified in the context of LTI systems.

The model reduction problem as formulated in controls is not immediately analogous to the parameter reduction problem addressed by MBAM.
For controls, ``model reduction'' refers to reducing the dynamic order of a system.
The goal is to approximate a system in the form of Eq.~\ref{eq:LTI}, with another LTI system with a smaller state space (i.e., $\tilde{x}(t) \in \mathbb{R}^{n'}$ where $n' < n$).

Balanced truncation uses the concepts of observability and controllability to identify accurate approximations.
For LTI systems, observability and controllability are each quantified by a Gramian matrix, $W_O$ and $W_C$ respectively.  
The Gramians are symmetric, positive definite matrices that act as metrics quantifying the observability and controllability of vectors in the state space.
A fundamental theorem for LTI systems shows that for an appropriate change of basis in the state space, the resulting observability and controllability Gramians are equal and diagonal.
In this so-called ``balanced representation,'' the vectors in the state space are equally controllable and observable.
\emph{Balanced truncation} is a reduction strategy in which an LTI system is first transformed into its balanced representation and the least observable/controllable states are truncated (removed) from the system.

To relate state-space reduction methods of control theory to MBAM, we parameterize the family of LTI systems of fixed dynamical order $n$.
We generated data by choosing many potential inputs $u(t)$ and evaluating the resulting time series $y(t)$.
The resulting model manifold is sloppy, and the sloppy \emph{parameters} are those components in the $A$, $B$, and $C$ matrices that determine the dynamics of the least observable/controllable \emph{states}.
When we apply MBAM to this system, we find the manifold boundary corresponding to the limit that the unobservable and uncontrollable states are constant decouple from the observed variables, $y(t)$.
This limit is equivalent to the balanced truncation method advocated by the controls community.

Balanced truncation has two major limitations.
First, it is difficult to interpret because physically meaningful states are mixed together in the balanced representation.
It is also difficult to generalize to nonlinear systems.
However, the information geometry techniques underlying sloppy model analysis naturally accommodate these regimes.
Thus, MBAM is a natural generalization of principles of balanced truncation to structure-preserving and nonlinear model reduction of interest to the control community.

\subsection{Beyond Manifold Boundaries}
\label{sec:beyondmbam}

\mkt{As presented here, MBAM has some limitations and raises some fundamental questions about the geometric and topological properties of multi-parameter models.
Here we discuss the relationship between MBAM and more general statistical theory.
We also highlight some interesting formal considerations related to the parameter singularities that are beyond the scope of this review.}

\mkt{In statistics, simplified models that are special cases of a more general model are called nested models.
Nested models are important in many statistical applications, especially hypothesis testing.
The boundary models that we describe here are a specific type of nested models.
But more generally, sometimes useful reduced models may also be found at non extreme values.
For example, in a spin system with parameters quantifying couplings (e.g., nearest neighbor, next nearest neighbor, etc.) that can be either positive or negative, a natural reduced model is to fix these couplings to zero rather than to extreme values.
Future work could adapt the MBAM to identify other types of useful reduced models that are not on the boundary.} \bbm{Whether there are geometric ways to identify these other nested models 
that don't make explicit reference to their having one less parameter remains an open question.}

\mkt{Taking parameters to extreme values is a common reduction strategy in science because the reduced model can often be evaluated in a different way.
For example, in a singularly perturbed dynamical system, an ordinary differential equation transitions into a differential-algebraic equation\cite{petzold1982differential-algebraic}.
In general, the singular nature of these limits makes them especially relevant for inter-theoretic reasoning in mathematical physics as in Figure \ref{fig:newtoneinsteinschrodinger}.
More broadly, however, model manifolds may have singular models that are not strictly boundaries, such as a the tip of a cone.
These singularities often correspond to useful reduced models.
More work is needed to understand the geometric and topological properties of these generalized singularities and develop reduction algorithms to efficiently find them.}


\section{Bayesian priors which \bbm{perform well for sloppy models}}
\label{sec:Priors}

\newcommand{\pmark}{p_\sharp}

A parameterized model $p(x|\theta)$ represents predictions for the
likelihood of seeing data $x$ in some experiment, given parameter
values $\theta$. The simplest question in Bayesian analysis is the
reverse: after seeing data $x$, can we infer the parameters? Bayes'
rule gives us not a single answer, but a distribution, proportional
to a {\em prior distribution} $p(\theta)$:
\[
p(\theta|x)=\frac{p(x|\theta)p(\theta)}{p(x)}
\]
where the denominator is simply a normalization constant. This result
is just mathematics, but deciding how to choose a prior (and exactly
what this means) has long been a delicate issue~\mca{\cite{kass1996selection}}. 
In some contexts we may have detailed pre-existing knowledge, and if we express this as a relatively narrow distribution $p(\theta)$, then Bayes' rule represents a small update.
But when we know little, we should express our ignorance as a broad distribution, \mca{some kind of} {\em uninformative prior}.
What concerns should guide us \mca{in formalizing this idea}, and how much do they matter?

One requirement is that put forward by Jeffreys: All results should be reparameterization invariant, and hence priors must transform like a density. 
For example, they should not depend on whether we choose to parameterize a model of exponential decay (like Eq.~\ref{eq:trivialModel}) with decay rates or with their inverses, lifetimes.
This would not be true for a flat prior \bbm{(meaning equal weight in parameter space)}, but it is true for Jeffreys prior, which is simply the volume form arising from the Fisher metric, up to normalization~\cite{jeffreys1946invariant}:%
\footnote{
  Despite bearing his name, this wasn't what he favored using in all situations.
  He argued for special treatment of location parameters, meaning directions along which the entire system is exactly symmetric, omitting them from the dimensions over which this volume form is calculated.
  The models considered here have no such parameters.
}
\begin{equation}
\label{eq:JeffreysPrior}
p_{J}(\theta)\propto\sqrt{\det g_{\mu\nu}(\theta)}.
\end{equation}
This volume measure encodes a continuum view of the manifold, in which it is infinitely divisible.
There is no notion that some parameter directions could be too short (too sloppy) to measure: Scaling the metric $g_{\mu\nu} \to M g_{\mu\nu}$ has no effect at all on $p_{J}(\theta)$ since $M$ is absorbed by normalization.
This scaling of distances by $\sqrt M$ is precisely the effect of repeating our experiment $M$ times, $p(x|\theta) = \prod_{i=1}^M p(x_i|\theta)$, and with enough repetitions, we can observe effects previously lost in the noise.
This continuum view makes $p_{J}(\theta)$ well-adapted for models in the limit of infinitely much data, but as we argued above, the multiparameter models appearing in science are very far from this limit.
Here, we argue that using it for such models can lead to extreme bias in relevant results.
Geometrically, the problem is that it places all of its weight \bbm{in regions where the hyperribbon is thickest in sloppy and thus unmeasurable directions.} 

A better behavior on a hyper-ribbon would be to place weight only \bbm{according to thickness along measurable, stiff directions, perhaps by being non-zero only} on appropriate edges of the manifold (Section~\ref{subsec:MBAM}).
A prior that did
this, schematically $p(\theta)=p_{\parallel}(\theta)\prod_{\mu\in\perp}\delta(\theta_{\mu})$,
would encode a lower-dimensional effective model. This section describes
some priors with this preference for edges. They open a view of model selection
as part of prior selection. 

\subsection{Maximizing mutual information}
\label{sec:OptPrior}

The natural measure of how much information we learn from an experiment
is the average reduction in the entropy over parameters $\theta$ upon seeing
data $x$, which is the mutual information~\cite{Lindley:1956bj}:
\[
I(X;\Theta)=S(\Theta)-S(\Theta|X)= \sum_x \int\negthickspace d\theta\, p(x|\theta)\,p(\theta)\log\frac{p(x|\theta)}{p(x)}.
\]
This information depends on the prior. 
Maximizing it may be thought of as choosing a minimally informative prior,
\mca{and this choice is one way to formalize what the notion of ``uninformative'' means.}\footnote{\mca{
Minimizing instead of maximizing, notice that a prior with all of its weight at any one point has mutual information $I(X;\Theta)=0$, the lowest possible value.
Such a prior represents complete certainty, and it seems fair to call it maximally} \bbm{biased}.
} The idea of \mca{choosing a prior by maximizing this information} was studied by
Bernando, but always in the limit \bbm{of} $M\to\infty$, infinitely many
repetitions or infinitely much data, where the resulting priors are called reference
priors~\cite{Bernardo:1979uq}. In this limit they approach Jeffreys prior
(under mild conditions), and this idea may be seen as an alternative motivation
for using Jeffreys prior~\cite{clarke1994jeffreys}.
But away from this limit, the resulting priors behave very differently,
without losing reparameterization invariance. And, in particular, they have
model selection properties which have been overlooked in the statistics
literature.

\begin{figure}
\centering \includegraphics[width=0.75\textwidth]{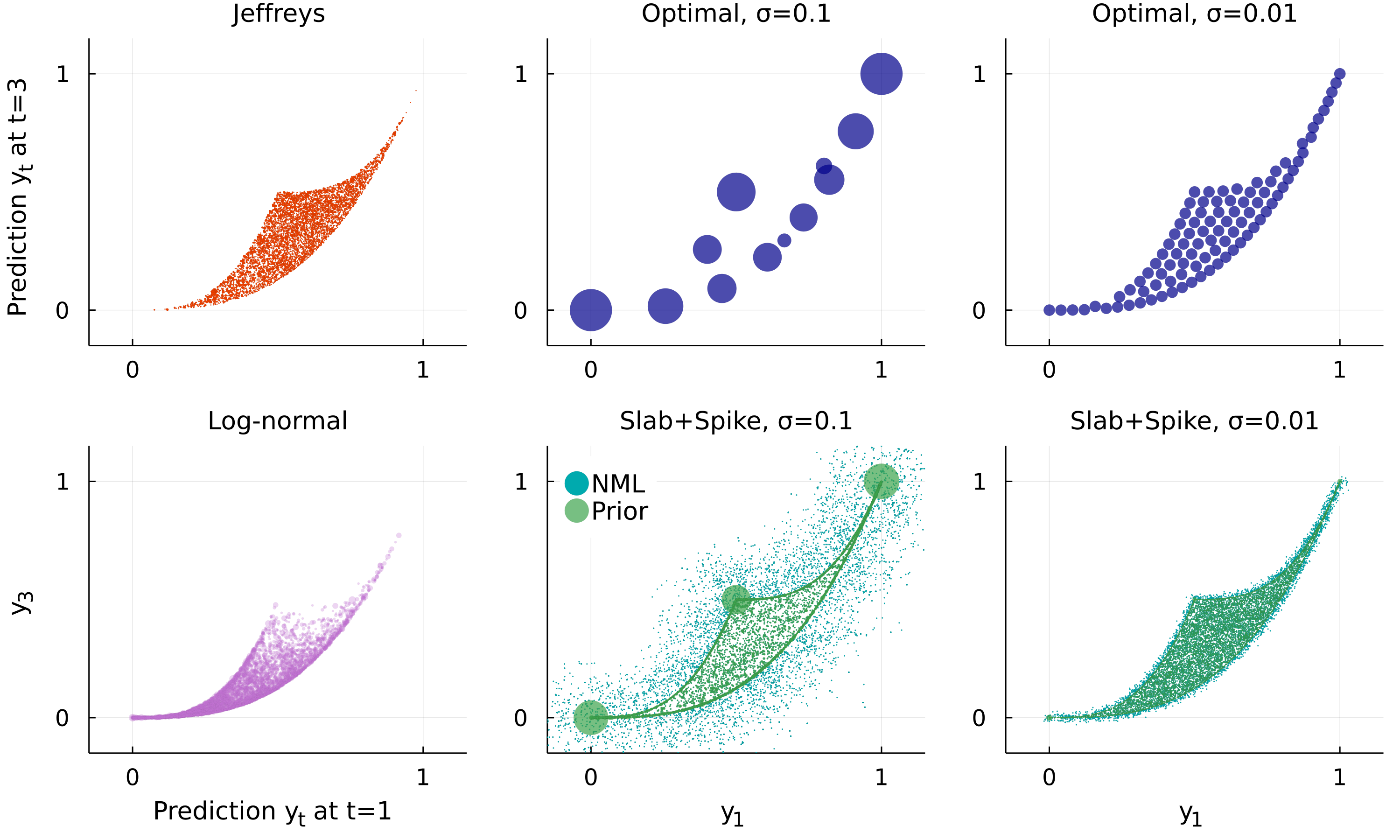}

\caption{\footnotesize Various priors for $D=2$ sum-of-exponentials models.
While Jeffreys prior (top left) is independent of $\sigma$, the discrete optimal prior $p_\star(\theta)$
(upper middle and right) has weight on zero- and one-dimensional edges at large noise, but adjusts to have 
weight in the two-dimensional interior as the noise is decreased.
\mca{The plots use as co-ordinates $y_t(\theta)$ for $t=1,3$, which is an isometric embedding onto the paper.}

The \mca{adaptive slab-and-spike} prior $\pmark(\theta)$ (lower right, discussed in Section~\ref{sec:MarkPrior}) also depends on the noise level.  At large $\sigma$, $p_\mathrm{NML}(x)$ has most of its weight outside the model manifold, hence $\pmark(\theta)$ has most of its weight on the edges (and uniform inside). At small $\sigma$, it approaches Jeffreys prior.
\label{fig:Discrete-optimal-priors-2D}}
\end{figure}

Priors which maximize mutual information were investigated in~\cite{Mattingly:2017uao}, which showed that 
they have a preference for edges, and hence tend to select simpler subspaces.
In fact they are almost always discrete, with weight on $K$ atoms (points in parameter space)~\mca{\cite{Farber:1967us, smith1971information, berger1989priors, Sims:2003cq}}:
\begin{equation}
\label{eq:prior-sum-atoms}
p_{\star}(\theta)=\mathop{\mathrm{argmax}}_{p(\theta)}I(X;\Theta)=\sum_{a=1}^{K}\lambda_{a}\delta(\theta-\theta_{a}).
\end{equation}
Along a short dimension (say $\lesssim1$ standard deviations) the
atoms will lie only at the boundaries. But along a long dimension
($\gg1$ standard deviations) there will be many atoms, approaching
a continuum. This turns out to be precisely the behavior we want on
a hyperribbon.

To illustrate this behavior, consider the sum-of-exponentials
model Eq.~\ref{eq:trivialModel}, used in Fig.~\ref{fig:cartoonManifold}, 
except with more parameters (and normalized to one at $t=0$):
\begin{equation}
x_{t}\sim y_{t}(\theta)+\mathcal{N}(0,\sigma^{2}),\qquad y_{t}(\theta)=\sum_{\mu=1}^{D}\frac{1}{D}e^{-tk_{\mu}(\theta)},\qquad0\leq k_{\mu}(\theta)\leq\infty.
\end{equation}
Fig.~\ref{fig:Discrete-optimal-priors-2D} shows $p_{\star}(\theta)$
for a model with $D=2$ parameters at various values of $\sigma$,
drawn in $y_{t}$ space for times $t=1,3$. 
\mca{(This choice of co-ordinates gives an isometric embedding: Euclidean distance on the paper is proportional to the Fisher metric.)}
At very large $\sigma$
there are just two atoms at the extreme values (i.e the zero-dimensional
boundaries). At very small $\sigma$, the data-rich regime, there
are many points almost tiling the space. And in the limit $\sigma\to0$
(equivalent to $M\to\infty$ repetitions) this approaches Jeffreys
prior, which is flat in $y$.
But it is the intermediate values which represent the more typical
hyper-ribbon situation, in which some dimensions are data-rich, and
some are data-starved.

\begin{figure}
\centering \includegraphics[width=1\textwidth]{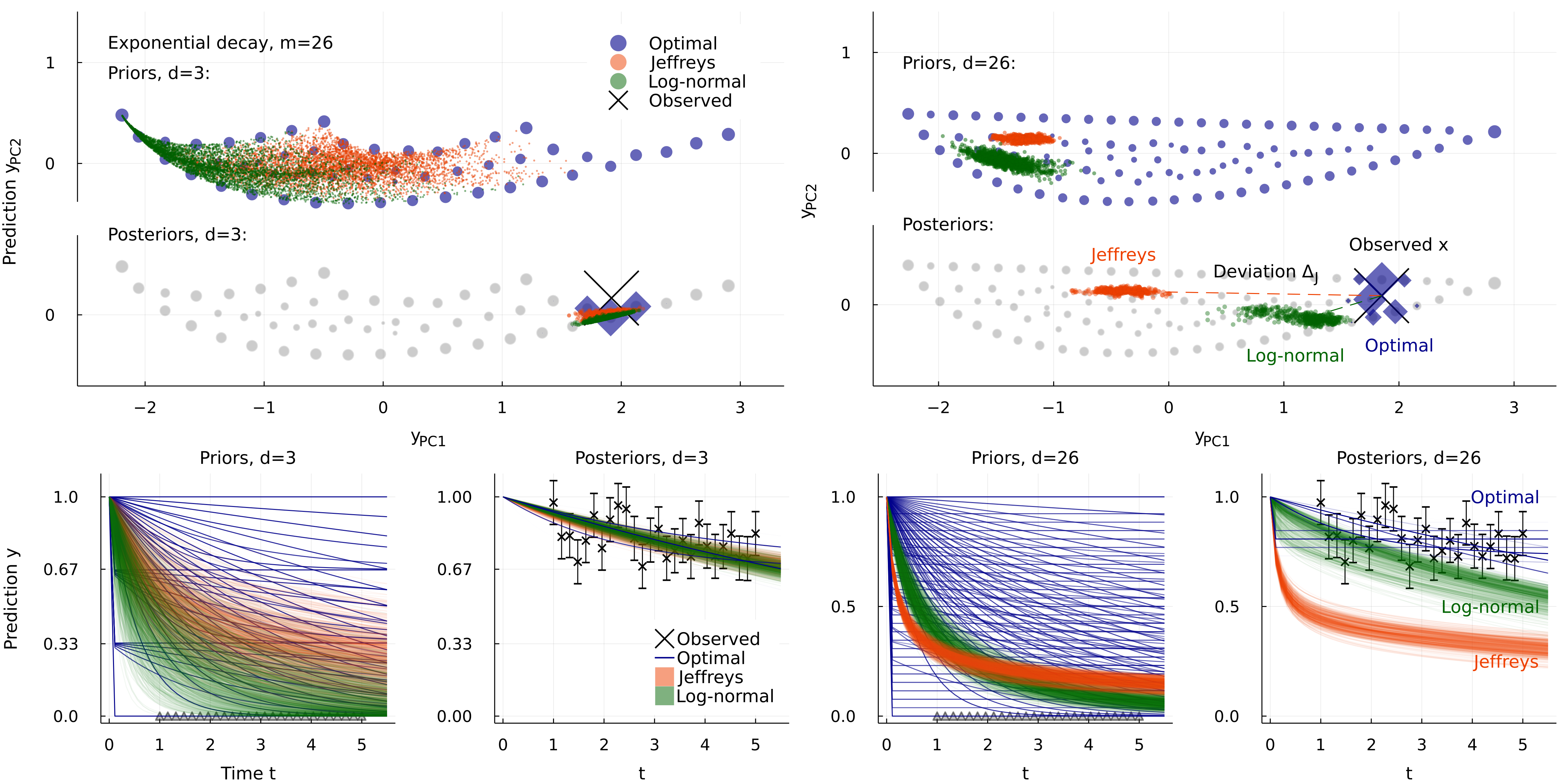}

\caption{\footnotesize Comparison of maximum-information and Jeffreys priors,
for $D=3$ (above) and $D=26$ (below) sum of exponentials models.
Above, the priors in prediction space $y$, showing how Jeffreys concentrates weight
in the center of the high-dimensional volume, while the points of the discrete 
optimal prior $p_\star(\theta)$ remain spread out. Lower left, the priors drawn
as time-courses $y_{t}$. Lower right, time-course plots of the posteriors,
showing the large bias of Jeffreys prior. The same observed data $x\in\mathbb{R}^{26}$
is used for both models, and also shown in the left panels; the error
bars indicate the scale of the noise, which can be compared to the
spacing of discrete points in the middle panels. Note how badly
Jeffreys prior distorts the best-fit posterior parameters (upper right) and
the fits to the data (lower right) for $D=26$.
\mca{(Figure taken from \cite{Abbott-2020-WIP}.)}
\label{fig:Comparison-Jeffreys-26D}}
\end{figure}

The idea of a prior which depends on the number of repetitions (or
equivalently on the noise level $\sigma$) breaks the dictum ``yesterday's posterior is tomorrow's prior''
for Bayesian updating.
\mca{However, any prior which depends on the experiment to be performed (including Jeffreys) already breaks this rule.
If tomorrow's experiment is upgraded to have less noise in some components, \bbm{even if it otherwise explores the same model}, then the combination requires a new prior,
thus the update from today to tomorrow \bbm{is different from a simple application of Bayes rule.  For Jeffreys prior there is an exception for repetitions of the same experiment.
}
But, as we demonstrate below, this invariance to repetition means that Jeffreys performs very badly for hyper-ribbon model manifolds, sloppy models, in which we cannot fix all parameters \cite{Abbott-2020-WIP}.}
Indeed, the distinction between a parameter we can fix and
one we cannot is precisely that we do not yet have enough data to fix
the latter.

\mca{The idea of accepting a discrete prior is also initially surprising to classical intuition.
What if the truth were between two points of weight, would we eventually have poor accuracy?
Three replies are possible.
First, ``eventually'' here implies taking much more data, and further updating the posterior, which is ruled out by the prior's dependence on the number of repetitions: The appropriate prior is the one for all the data.
Second, along a relevant parameter direction, discretization is arguably no worse than truncating some measured $\theta$ to a sensible number of digits, for given error bars.\footnote{
In fact the optimal prior is closer to continuous than this sentence suggests.
The number of atoms $K$ grows slightly faster than the length of a dimension
$L$, by a novel power law~\cite{Abbott:2017oqw}.
}
And third, along an irrelevant direction, placing all weight at an extreme value is precisely what effective theories do. For example, QED assumes that the coupling constants of all forces except electromagnetism are exactly zero, which we know to be false in reality. But this false assumption is irrelevant to the experiments being described, and allows predictions of very high accuracy \cite{aoyama2019atoms}. 
}

Jeffreys prior is, as we mentioned, the limit $M\to\infty$ of the
maximum-information prior $p_{\star}(\theta)$. So according to our
arguments, it's the correct thing to use in the regime when all parameters
are well-constrained, when all are in the data-rich regime. How badly
wrong is it when we are not in this regime?
The eigenvalues of the metric tensor in sloppy, multiparameter models span
an enormous range, so the square root of the determinant giving Jeffreys
prior (Eq.~\ref{eq:JeffreysPrior}) is often enormously small. Worse, it
varies often by many orders of magnitude even for small changes in parameters.
Uninformative priors are supposed to be broadly agnostic about parameter 
choices, and Jeffreys prior fails in this regard.

Thanks to progress in numerical techniques for finding $p_{\star}(\theta)$~\cite{Abbott-2020-WIP, huber2008entropy} we can now illustrate this extreme weighting.
Fig.~\ref{fig:Comparison-Jeffreys-26D} shows the same sum-of-exponentials
model, but now comparing a model with $D=3$ parameters to one with $D=26$ parameters.
Both describe the same data, measuring decays at $M=26$ different times in $1\leq t_i \leq5$, each with noise $\sigma=0.1$. 
The discrete priors for these two models are fairly similar, but the concentration
of weight in Jeffreys prior is much stronger in the higher-dimensional
model. This is most clearly seen in the posterior, after observing
a particular $x$ as shown. In $D=3$ Jeffreys performs well, but
in $D=26$ its posterior is centered 20 standard deviations away from
the data. This is not a subtle error.

How many repetitions would be required in order for Jeffreys prior
to be justified here? The smallest eigenvalue of the FIM is about
$10^{-50}$ here. To make this order 1, we need more repetitions
than there are seconds in the age of the universe \cite{Abbott-2020-WIP}. While this is obviously
a toy model, tens of irrelevant parameters, and astronomical timescales
to measure them all, are also found in more realistic models 
(Figs.~\ref{fig:sloppyEigs} and~\ref{fig:sloppyLengths}).

\begin{figure}
\centering \includegraphics[width=0.6\textwidth]{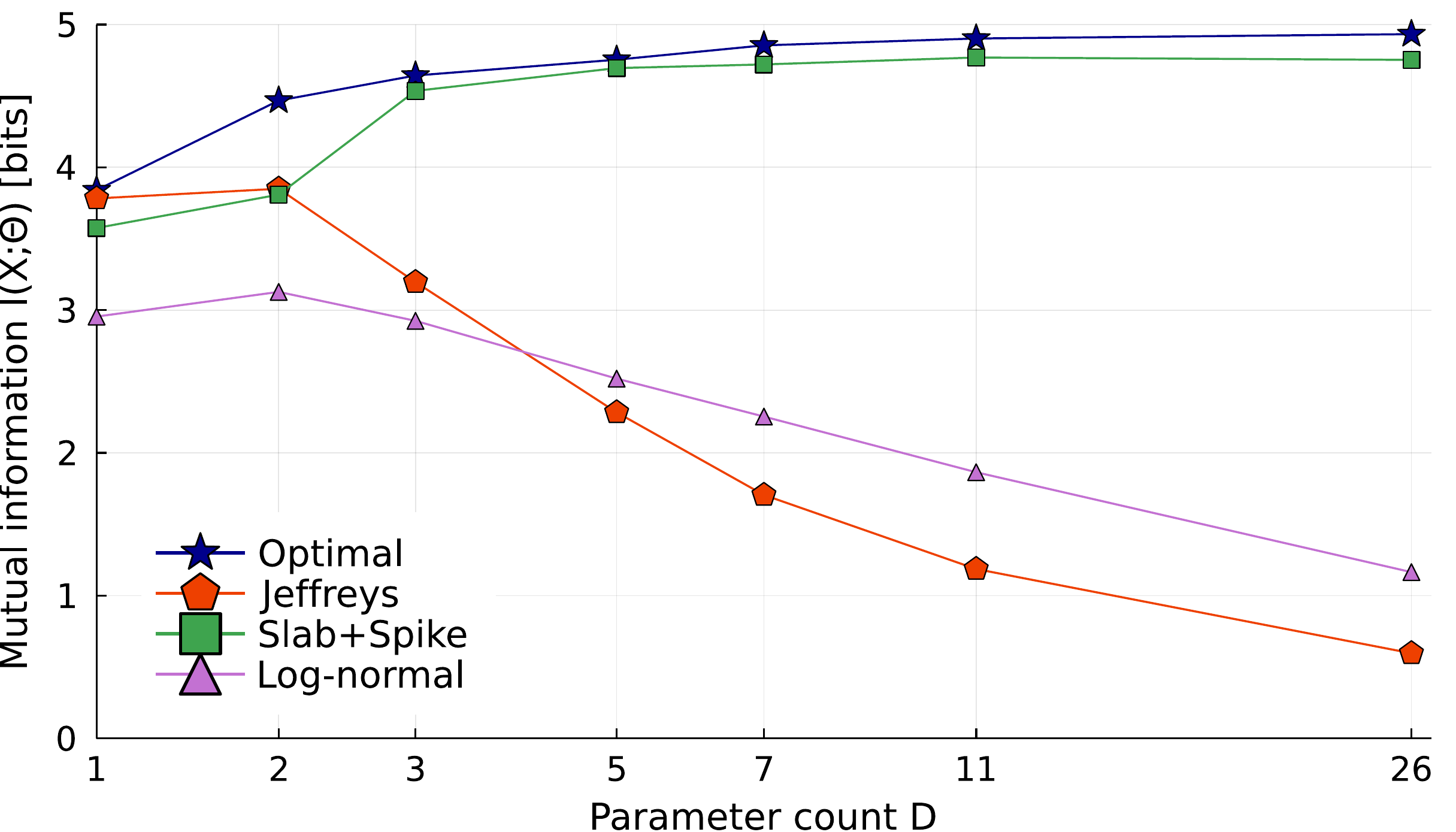}\caption{\footnotesize 
Mutual information $I(X;\Theta)$ captured when using various priors for sum-of-exponentials model, 
against the number of model parameters $D$. Jeffreys and the optimal prior for $D=3$ and $D=26$ are
drawn in Fig.~\ref{fig:Comparison-Jeffreys-26D} above, while the projected maximum likelihood prior
is drawn in Fig.~\ref{fig:Mark-Jeffreys-26D} below.
Notice how close the projected prior is to the optimal prior in high dimensions,
and how poorly Jeffrey's and the log-normal priors perform by this measure.
\label{fig:MI-comparison}}
\end{figure}

Another way to compare these priors is by how much information we
will learn. The scores (expressed in bits) for the priors in 
Fig.~\ref{fig:Comparison-Jeffreys-26D} are shown in 
Fig.~\ref{fig:MI-comparison}. 
Notice that adding additional parameters almost always increases $p_{\star}(\theta)$,
but we observe that the extreme concentration of weight gives Jeffreys
prior a much lower score in the high-dimensional model. Even without
being able to find $p_{\star}(\theta)$ exactly, this measure offers
a tool for model selection. But unlike traditional approaches to model
selection, this takes place before seeing data $x$. It depends only
on the resolution of the experiment, not its outcome.

\subsection{\mca{Adaptive slab-and-spike priors}}  

\label{sec:MarkPrior}

Finding $p_{\star}(\theta)$ is not always easy, as $I(X;\Theta)$
involves both an integral over $X$ and a sum over the atoms: 
The $K$ points of weight all interact with each other, via the data they generate.
We present here another prior which shares many of the virtues (and avoids
Jeffreys' vices), while being easier to draw samples from.

\begin{figure}
\centering \includegraphics[width=1.0\textwidth]{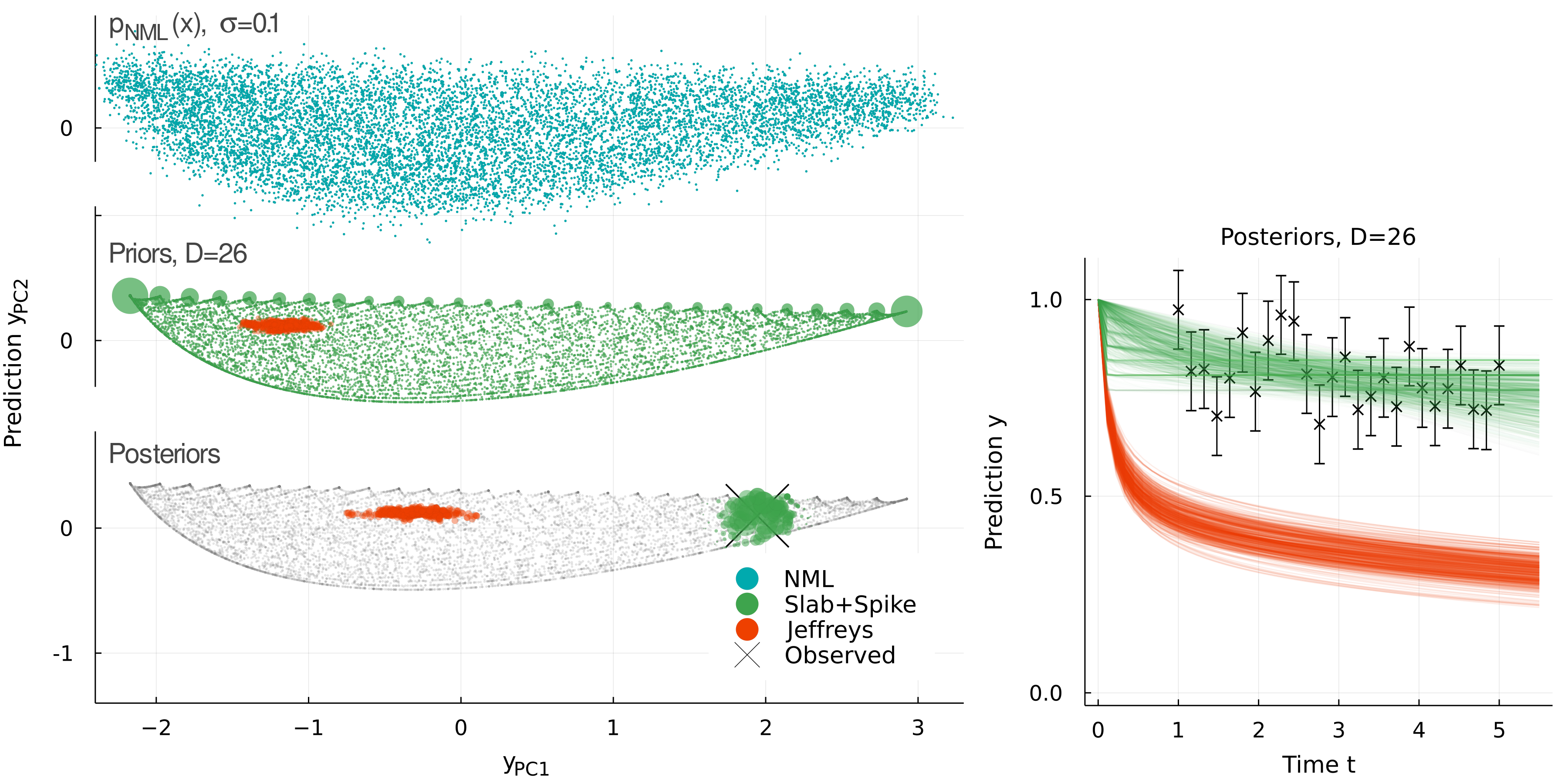}
\caption{\footnotesize 
Behavior of the \mca{adaptive slab-and-spike} prior $\pmark(\theta)$, 
for the same $D=26$ system as in Fig.~\ref{fig:Comparison-Jeffreys-26D}, 
with Jeffreys prior repeated for comparison.
The distribution $p{_\mathrm{NML}}(x)$ which when projected onto the model manifold gives the prior $\pmark(\theta)$ is drawn shifted upwards on the plot.
As before, the posteriors in $y$-space are drawn shifted down on the plot. 
The panel on the right shows the posteriors as functions of time. 
\label{fig:Mark-Jeffreys-26D}}
\end{figure}

To derive this, consider first the following distribution on data space $X$, called a normalized maximum likelihood:%
  \footnote{This is also used, independently, in a \mca{non-Bayesian} approach to model selection
  by minimum description length~\cite{myung2006model, grunwald2019minimum}.
  In that context there is no measure on $\Theta$, although a measure on $X$ is needed to calculate the normalization $Z$. The normalization $Z$ is not crucial, and indeed can be infinite (an {\em improper} prior); such a prior can still be sampled with Monte Carlo (see below).}
\begin{equation}
p_{\mathrm{NML}}(x)=\frac{\max_{\hat{\theta}}p(x|\hat{\theta})}{Z},\qquad Z=\int dx\:\max_{\hat{\theta}}p(x|\hat{\theta}).
\end{equation}
For models with Gaussian noise, this describes a cloud of points filling and surrounding the embedded model manifold \mca{$\mathcal{Y} \subset \mathbb{R}^M$} (Figs.~\ref{fig:Discrete-optimal-priors-2D} and~\ref{fig:Mark-Jeffreys-26D}), with a range given by
the measurement error in the predictions. We then define a prior
distribution on parameter space $\Theta$, by simply projecting
the weight of $p_{\mathrm{NML}}(x)$ at each $x$ to the corresponding
maximum likelihood point $\hat{\theta}$. We may write this as
\begin{equation}
\pmark(\theta)=\int dx\;p_{\mathrm{NML}}(x)\:\delta\big(\theta-\mathop{\mathrm{argmax}}_{\hat{\theta}}p(x|\hat{\theta})\big).
\end{equation}
This prior \mca{has full support, but} places extra weight on points on the boundaries of the model manifold,
and more weight on lower-dimensional boundaries (such as fold lines, and corners),
\mca{thus generalizing the idea of a slab-and-spike prior \cite{mitchell1988bayesian} to many dimensions and curved manifolds}.
\mca{How much weight is on boundaries is chosen automatically, depending on the level of noise in the likelihood function.}

In practice, this prior can be computed efficiently by sampling $p_{\mathrm{NML}}(x)$ using Monte Carlo methods, and then 
recording for each accepted point $x$ the corresponding $\hat{\theta}$ to
give a sampling distribution $\pmark(\theta)$ in parameter space.
The Metropolis weight does not require the normalization factor $Z$.

Figures \ref{fig:Discrete-optimal-priors-2D} and~\ref{fig:Mark-Jeffreys-26D} show the result, in the same sum of exponentials models as above. 
For all models $x\sim y(\theta)+\mathcal{N}(0,\sigma^{2})$,
the embedding of the model manifold is smeared out by a distance $\sigma$
to make $p_{\mathrm{NML}}(x)$. This means that, unlike Jeffreys prior,
the difference between places where the manifold is thin compared
to $\sigma$, and places where it is very very thin, won't matter
at all. The smeared weight is pulled back to contribute to $\pmark(\theta)$
exactly at the closest edge. And lower-dimensional edges have more
weight; zero-dimensional points usually have finite weight.
Thus $\pmark(\theta)$ is as a generalized slab-and-spike prior,
with weight both in the volume and on edges. 

Like $p_{\star}(\theta)$, the degree of preference for edges depends
on the level of noise $\sigma$, or the number of repetitions. Along
irrelevant, sloppy, short, directions, \mca{almost} all the weight is on the edges.
Along relevant, stiff, long, directions, most of it \mca{fills the interior}.
As the noise is decreased, dimensions will smoothly pass from the
former category into the latter. We view this as a virtue, but it
does mean the prior $\pmark(\theta)$ is not invariant to repetition
of the experiment.

While it does not maximize the information $I(X;\Theta)$, the prior
$\pmark(\theta)$ usually does very well on this measure. Figure \ref{fig:MI-comparison}
above shows an estimate of this, compared to $p_{\star}(\theta)$
as well as Jeffreys prior, and a log-normal prior.

\section{Continuum limits in physics, and the renormalization group}
\label{sec:ParameterCompression}

Common to many areas of science are models in which key properties are emergent---not easily discernible from the properties of microscopic components.  In many cases, a microscopically motivated model might naturally contain a large number of parameters, while the emergent behavior can be described with a much more compact effective theory.  Conceptually, most parameter combinations of the underlying model are \textit{irrelevant} for their emergent behavior.  An effective theory can suffice provided it is sufficiently flexible to capture the smaller number of \textit{relevant} parameters in the microscopic description.

These ideas are most precise for models which are \textit{renormalizable}~\cite{Sethna06,Goldenfeld18}.  In renormalization group (RG) analysis, a microscopic model is coarse-grained by iteratively writing down an approximate description for the behavior of a subset of the full microscopic degrees of freedom.  For example, in real space RG under decimation, for a Hamiltonian which defines the probability distribution of spin states on a square lattice, a renormalized Hamiltonian should give the `effective Hamiltonian' for the subspace of lattice sites for which $i$ and $j$ are even.  Importantly, this Hamiltonian lives in the same parameter space as the microscopic one, though in general the renormalized Hamiltonian will have different parameters.  This coarsening, (together with a rescaling) defines a change of parameters, and is known as an RG step; the movement of parameters under repeated iterations is known as an RG flow.

Emergent behavior is seen by considering the long-time behavior of RG flow after which models sit nearby to RG fixed points. These fixed points define possible {\em universality classes} of coarse-grained behavior.  Phase transitions, and other emergent phenomena sit at nontrivial fixed points.
A linearization of the flow equations close to these non-trivial fixed points has a small number of directions which are unstable, while the rest are stable.  Each of these unstable directions is a relevant parameter---a small change in microscopic parameters will lead to a large change in the emergent behavior, seen after many rounds of coarse-graining.  Conversely, stable directions of the flow are irrelevant---moving along them in the microscopic model will have an exponentially decreasing effect on the parameters of the emergent description. The Ising critical point has two relevant parameters (magnetic field $h$ and reduced temperature, $t=(T-T_c)/T_c$), which are measured by the projection of the parameters onto the left eigenvectors of the linearization of RG dynamics near the Ising fixed point. 

\subsection{Coarse-graining and the Fisher metric}

Because many models are microscopically defined, but are mostly observed through their emergent properties we sought to understand how the model manifold of microscopic parameters contracts when observables are coarse-grained.  Rather than coarse-graining the parameters to directly build an effective description as is done in a usual RG treatment, we instead wanted to understand how coarse-graining changes one's ability to measure microscopic parameters~\cite{Machta:2013ga,Raju:2017ty}.  To do this we consider a continuous coarse-graining where as the smallest length scale $l=l_0\exp{b}$ changes the parameters change according to 
$d\theta^\mu / db =\beta^{\mu}$ where $\vec{\beta}$ are the beta functions that define the flow.  Infinitesimal distances after such a flow can be calculated by flowing the parameters under the usual RG, and then measuring their distance in the coarse grained model, where the system size is smaller.  This change in the microscopic parameter space metric can be summarized by the following equation~\cite{Raju:2017ty}:
\begin{equation}
\label{eq:modifiedLie}
\frac{d g_{\mu \nu}}{db}= \underbrace{ \beta^{\alpha} \partial_\alpha g_{\mu \nu}}_\textrm{$g'$ along flow} + \underbrace{g_{\alpha \mu} \partial_{\nu} \beta^{\alpha} + g_{\alpha \nu} \partial_{\mu} \beta^{\alpha}}_\textrm{contraction of $\Delta \theta$}- \underbrace{D g_{\mu \nu}}_\textrm{Vol. contraction}.
\end{equation}
The first two terms constitute the Lie derivative of the metric under flow $\beta$.  The third term arises because the metric is extensive, proportional to the system size $L^D$, where $D$ is the dimensionality of the system.  


While this equation summarizes the local change in the metric, and as such the change in infinitesimal distances during coarse-graining, it remains to analyze how these distances change under typical RG flows.  For this analysis it is useful to use the fact that the metric can be written as the second derivative of the free energy (in units where $\beta=1$), $g_{\mu\nu}=\frac{\partial}{\partial \theta^\mu} \frac{\partial}{\partial \theta^\nu} F$.  In renormalization group analysis it is useful to separate the free energy into a singular and an analytic piece, $F=F_a+F_s$, where the singular part is non-analytic as a function of its variables near the RG fixed point~\cite{Goldenfeld18}. Near the RG fixed point the metric, like the free energy, can be also divided into singular and analytic pieces, $g=g^s+g^a$.  By assumption, the singular contributions of the free energy are those that are preserved under RG flow---the three terms in Eq.~\ref{eq:modifiedLie} must precisely cancel. This means that along relevant directions---which are best measured through their influence on emergent properties, the microscopic metric does not contract.%
  \footnote{This behavior in prediction space is quite different from the
  flows in parameter space, where the relevant directions grow under the
  renormalization-group transformation.}
The preservation of the metric under flow can be used to constrain the form of the metric along relevant directions~\cite{Reevu15}.  However, these constraints only apply to the lower dimensional subset of parameters which are relevant in the usual RG treatment.

Irrelevant directions are instead measured through their influence on microscopic details and their metric is dominated by analytic contributions.  For these directions the first term is expected to be small and infinitesimal distances $ds$ contract according to
\begin{align}
\frac{d\log (ds^2)}{db} =-D+2y_i,
\end{align}
where $y_i$ is the corresponding (negative) RG exponent.

\begin{figure}
\centering \includegraphics[width=.7\textwidth]{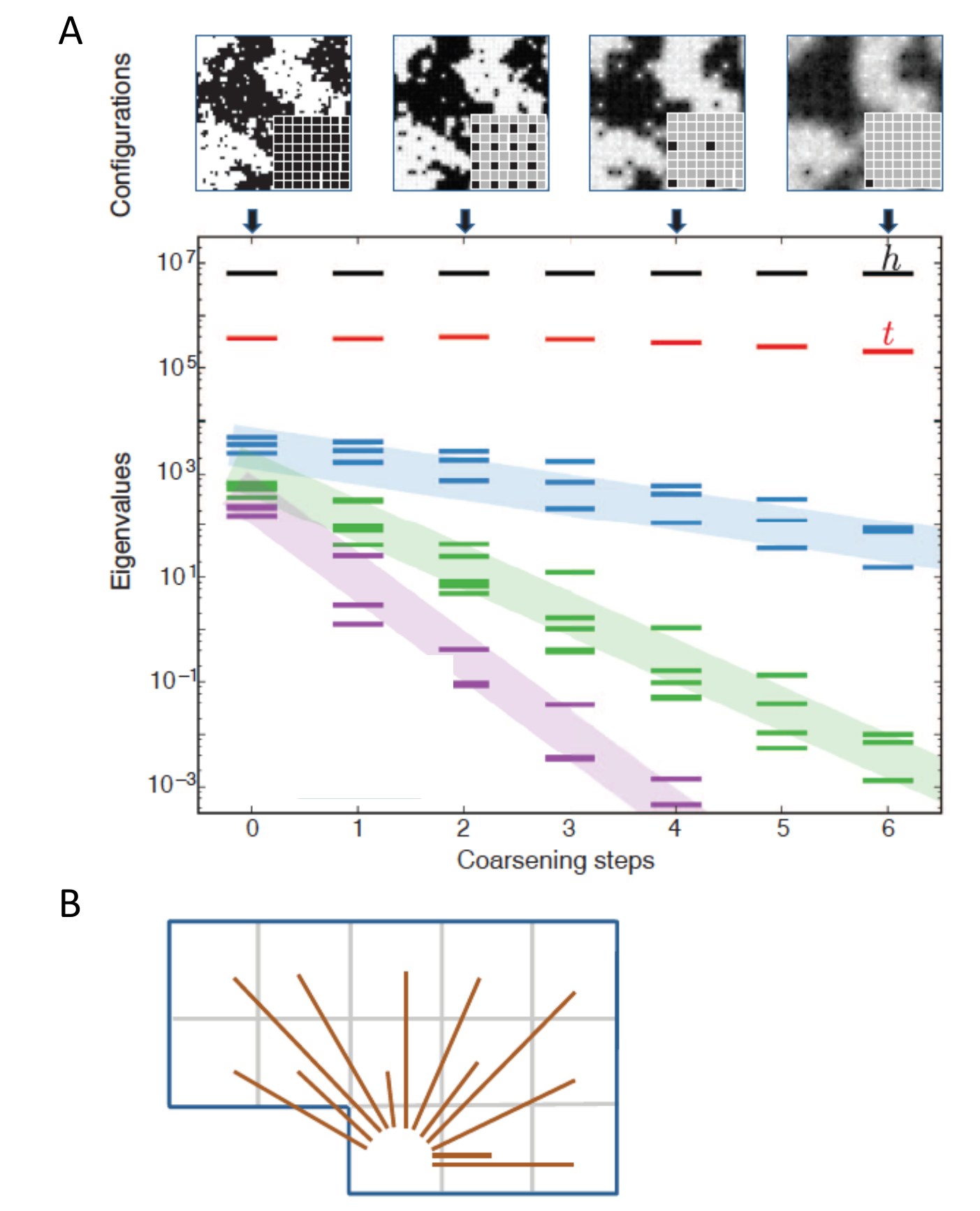}

\caption{\footnotesize (Adapted from~\cite{Machta:2013ga}).  (A)  FIM spectrum is plotted vs level of coarse-graining for a thirteen parameter `nearby neighbor' model where parameters describe translation invariant couplings between spins as shown in (B) as well as a magnetic field.  Without coarse-graining (0 steps), by observing entire configurations, (black in inset) relevant parameter combinations are easy to infer (red and black dashed lines) and other combinations have a typical scale.  As coarse-graining proceeds (black squares in inset remain observable) relevant parameters remain easy to infer, but irrelevant ones contract with their RG exponent, slope $(-D+2y_i)/2 \log(2)$ since each coarse-graining step contracts lengths by a factor of $\sqrt{2}$.  Here blue eigenvalues have $y_i=0$, green have $y_i=-1$ and purple have $y_i=-2$  
\label{fig:IsingCoarseGrain}}

\end{figure}

We have verified this prediction numerically for a two dimensional Ising model, where we have added couplings between non-nearest-neighbor spins (See Fig.~\ref{fig:IsingCoarseGrain} and Ref.~\cite{Machta:2013ga}).  We coarse-grain the system using simple decimation, only observing the red sites on a checkerboard, and so on recursively.  In the Ising model these results have an intuitive interpretation; the magnetic field and temperature can be measured easily without seeing individual spins as they control the net magnetization and the correlation length.  However, other parameter combinations can only be measured by examining microscopic configurations.  

\subsection{Global model manifold flows}
\begin{figure}
\centering \includegraphics[width=.5\textwidth]{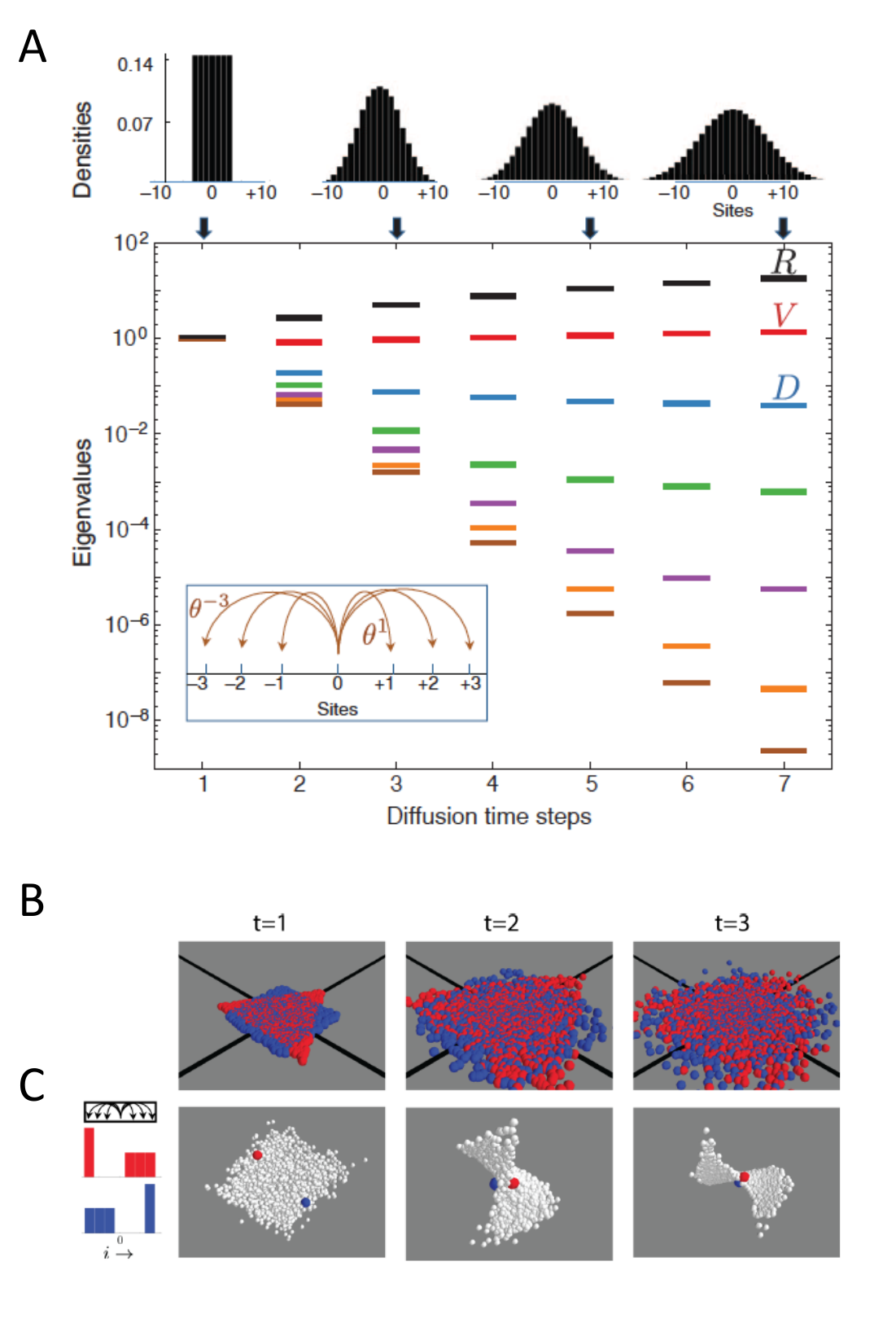}

\caption{\footnotesize Coarse graining a model of hopping yields a lower dimensional diffusion equation.  (A)  The Fisher Information's eigenvalues for a model where particles hop according to a seven parameter kernel in each time step, with each parameter describing the rate of hopping $\mu$ sites on a discrete lattice in a discrete time step.  After a single time step the metric is the identity, but if the density is observed only after taking multiple steps then the distribution becomes sloppy---the largest eigenvalue controls $R$, non-conservation of particle number, the next largest the drift, $V$, followed by the diffusion constant, $D$, and a hierarchy of irrelevant parameters that control higher cumulants of the density.  As time evolves the central limit theorem dictates that density (above) should approach a Gaussian, parameterized by the three largest numbers, $R$ (controlling the total number of particles), $V$ (controlling the mean of the distribution) and $D$ (controlling its width).  (B) This is shown pictorially in a model where blue particles hop according to a square kernel while red particles hop according to a triangular kernel.  After a single time step their distribution is not similar.  However, because their diffusion constant and drift velocity are matched, the particles' distributions become similar as time evolves.  (C) In the one-parameter example above, kernels (parameter sets) are drawn uniformly from the space that conserves particle number, and their model manifold's projection onto a three parameter subspace is shown at three time points. After a single time point the model manifold is not a hyper-ribbon, but is extended roughly equally in all directions. Two parameter values, marked in red and blue, are separated substantially even though they have the same values of $V$ and $D$.  After coarsening the model manifold collapses towards a two-dimensional surface which could be parameterized by drift and diffusion alone, and the two marked points become close to each other. 
(Adapted from~\cite{Machta:2013ga,Transtrum:2015hm,Raju:2017ty}.)
\label{fig:DiffCoarseGrain}}

\end{figure}

While this metric centered approach provides information about how the local structure of the manifold contracts under coarsening, it does not necessarily provide information about the global structure of the manifold.  While we have fewer analytical approaches, we can numerically investigate how an entire model manifold contracts under coarse-graining.  We can numerically investigate the diffusion equation, where a simple renormalization group treatment yields a single relevant parameter (the drift, $v$) a single marginal parameter (the diffusion constant $D$) and an infinite number of irrelevant parameters which control higher cumulants of the distribution~\cite{Raju:2017ty,Machta:2013ga}.  Numerically we investigate a hopping model, where parameters $\theta^\mu$ describe the probability that a single particle moves a distance $\mu$ in a discrete time step (see Fig.~\ref{fig:DiffCoarseGrain}A).  We consider a seven parameter model when particles are allowed to move up to three spaces to the left and right; particle number is not necessarily conserved, but there is a six dimensional subspace which conserves particle number (the subspace where $\sum_\mu \theta^\mu=1$).

The metric can be measured at a single point in parameter space by observing the particle density after some number of time steps.  Its eigenvalue distribution is shown in Fig.~\ref{fig:DiffCoarseGrain}A for the parameters where $\theta^\mu=1/7$.  In the upper panel of Fig.~\ref{fig:DiffCoarseGrain}A, the particle distribution is shown after $t$ time steps.  Initially the distribution is given by the parameters, $\rho_i \propto \theta^i$, and each parameter can be measured independently, yielding a metric proportional to the identity, which is not sloppy.  After sufficient coarse-graining steps, the distribution approaches a Gaussian as required by the central limit theorem, characterized by a velocity and diffusion constant, and, if particle number is not conserved, by a rate of particle creation/annihilation ($R$). In this limit, the spectrum of the Fisher Information becomes sloppy, with the stiffest eigenvector changing $R$, the next stiffest changing $V$, then $D$, and with a hierarchy of irrelevant parameters which manifest only through higher cumulants in the distribution.

The non-local contraction of distances can be seen qualitatively by looking at two (two-dimensional) diffusion kernels, one a triangle, one a square, whose drift is both zero and which have the same diffusion constant (Red and blue points, respectively in Fig.~\ref{fig:DiffCoarseGrain}B).  After a single time step these two kernels yield very clearly distinguishable particle distributions.  However, after multiple time steps the two distributions become similar, approaching the same Gaussian.  We numerically investigated the seven parameter model, by taking kernels uniformly from the 6 parameter space of particle conserving kernels, and plotting them in the first three principle components of the space of particle densities.  After a single time step the points fall in a uniform three dimensional blob - the corresponding model manifold does not have a hyper-ribbon structure.  The red and blue points, which each have zero drift and the same diffusion constant, are far apart in this model manifold.  However, as time evolves the three dimensional blob collapses onto a two dimensional structure parameterized by a drift and diffusion constant.  As this limit is approached the red and blue points approach each other rapidly. 

In renormalizable models, after coarse-graining the model manifold contracts so that every point is close to a smaller dimensional surface which can often be described by a reduced model with just relevant parameters kept. This gives an information theoretic perspective for one of the central practical findings of the renormalization group-  that a simple `toy model' will often suffice to describe a microscopically complicated system.  Many systems are not renormalizable; however, they still often show a sloppy spectrum in their Fisher Information's eigenvalues and in the hyper-ribbon structure of their model manifolds.  These systems would often not be sloppy if they were observed at some suitable microscopic scale.  However, for macroscopic behavior, only a subset of their microscopic properties remain important, leading to local sloppiness and a global hyper-ribbon structure.  In that sense the sloppiness that arises in renormalizable systems may be qualitatively similar to that arising in other systems. One key difference is that renormalizable models have a self-similarity during coarse-graining, allowing metric flow to be written compactly in terms of parameter space point flow (as in Eq.~\ref{eq:modifiedLie} from Ref.~\cite{Raju:2017ty}, see Ref.~\cite{Strandkvist20} for an interesting perspective).

\section{Replicas and Minkowski embeddings for visualizing models}

\label{sec:MinkowskiSpace}

An essential component of working with model manifolds and hyperribbons is being able to map and visualize them. This is often achieved by sampling the manifold in a region of interest (see Section~\ref{sec:Priors}), and plotting the resulting data set. However, finding an accurate way of visualizing complex, high-dimensional data is a difficult task. It is often not possible to faithfully represent both distance and geometry with fewer dimensions than the original data, and so it would seem that visualizing model manifolds may also not be possible. For least-squared models, 
we have shown both empirically~(Fig.~\ref{fig:sloppyLengths}) and rigorously (Section~\ref{sec:HyperribbonBounds}) that the model predictions lie on flat hyperribbons, which makes Principal Component Analysis (PCA) visualization an effective tool for data formed by varying parameters in the model (e.g., Fig.~\ref{fig:Comparison-Jeffreys-26D}).

Other physical models do not so neatly conform to this approach. For instance, the canonical Ising model from statistical physics predicts the likelihood of a certain spin configuration given fields and bond strengths. This does not fit the structure of a least-squares model, since the probability distributions are not
Gaussian fluctuations about predicted data points, but instead a probability
distribution over a data set of spin configurations.
The $\Lambda$-CDM cosmological model of the early universe similarly provides a 
prediction for the likelihood of a given sky map: the 
probability density of finding a pattern of temperature and polarization fluctuations in the cosmic microwave background radiation. A new approach is therefore necessary to visualize these kinds of general probabilistic model manifolds, which we shall for simplicity refer to as probabilistic models in this section.

The task of visualizing high-dimensional data is well known in the data
science and machine learning communities, and has fueled a growing
number of visualization techniques to extract lower-dimensional
visualizations that faithfully capture important and relevant properties
of the manifold that are of interest to researchers. 
\jps{Examples include t-SNE~\cite{van2008visualizing},
diffusion maps~\cite{coifman2005geometric}, UMAP~\cite{mcinnes2018umap}, 
Isomap~\cite{tenenbaum2000global} and hyperbolic mapping~\cite{boguna2010sustaining}; there are many
reviews~\cite{izenman2012introduction} and software
environments~\cite[Section 2.2]{scikit-learn} to guide newcomers.}
In particular, all
methods must address in some way the \textit{curse of dimensionality},
where distance measures in ever increasing dimensional spaces lose
useful distinguishability, mainly because the volume increases so
rapidly that finite data sets become sparse and orthogonal. As a result,
all visualization methods must compromise in some way. Most techniques
accomplish this through a trade-off between global and local structure.
For instance, classical manifold learning methods, such as Principal
Component Analysis (PCA)~\cite{pearson1901liii,hotelling1933analysis}
and Multidimensional Scaling (MDS)~\cite{torgerson1952multidimensional}
preserve global properties at the expense of often obscuring local features.
Conversely, most modern nonlinear methods, such as t-distributed
stochastic network embedding (t-SNE)~\cite{van2008visualizing}, diffusion maps~\cite{coifman2005geometric}, and hyperbolic
mapping~\cite{boguna2010sustaining}, highlight local structures while only qualitatively maintaining
certain global features such as clusters.

There are similar problems for probabilistic models. Unlike least-squares, these models do not have a simple, finite-dimensional Euclidean space in which they can be easily embedded to reveal important features. If one wishes to preserve both local and global features, then a different compromise must be made. When visualizing collections of probability distributions, 
we have uncovered visualization methods which perform as well as PCA does for
least-squares models, by embedding in a pseudo-Riemannian space 
(\textit{i.e.} a space with both time-like and space-like components,
for which 3+1-dimensional Minkowski space is the canonical example). Two
particular choices are Intensive Principal Component Analysis (InPCA~\cite{Quinn:2019dc}) and Intensive Symmetrized Kullback-Leibler Embedding (isKL~\cite{Teoh:2020kp}.).

Using these methods, many physical models are represented as 
an isometric, low-dimensional hyperribbon embedding that
quantitatively preserves both global and local features. The time-like
coordinates with negative squared distance do make interpretation less intuitive, but often serve to illuminate important model features (Fig.~\ref{fig:isKLIsing}).
Both InPCA and isKL give hierarchical embeddings. However, isKL has the added advantage of producing a finite embedding for exponential families---although the data space for distributions can be infinite dimensional, isKL systematically extracts a space-like and a time-like component for each \textit{parameter} in the model itself, in a hierarchical manner that reveals the underlying hyperribbon features. 

\subsection{Why Euclidean embeddings are doomed}
\label{subsec:EuclidDoomed}

\jps{To understand why it is that embedding in a pseudo-Riemannian space is more natural for manifolds of probabilistic models, consider the failures of a natural embedding method, using the Hellinger distance 
\begin{equation}
\label{eq:Hel-defn}
d^2_\mathrm{Hel}[\theta,\tilde\theta] = 4 \sum_x \left(\sqrt{p(x|\theta)}-\sqrt{p(x,\tilde\theta)}\right)^2.
\end{equation}
Since probabilities are normalized, $1 = \sum_x p(x|\theta) = d^2_\mathrm{Hel}[\theta,\theta]/4$, the vectors $\mathbf{n}_x  = 2 \sqrt{p(x|\theta)}$ are points
on a hypersphere of radius two in a space whose dimension is the cardinality of the observations. The Hellinger distance is the Euclidean
distance between these points. As $\theta$ is varied, the points form a manifold
\begin{align}
\label{eq:HellVector}
y_x(\theta) = 2\sqrt{ p(\bx_i|\theta)} 
\end{align}
which can be visualized with PCA.  The metric between nearby points on this manifold is given by the Fisher information,%
  \footnote{Why do we take a square root? Nearby probability vectors
  $y_\bx(\theta) = p(\bx|\theta)$ have squared differences that do not correspond
  to the Fisher information distances. By taking the square root \jps{and multiplying the lengths by two}, we fix this: the Hellinger embedding is {\em isometric}.}
it is {\em isometric}.
The fact that it is isometric makes it a natural embedding,} in the sense that distance between points represents the distinguishability of the different probability distributions (\textit{e.g.} if two distributions have no overlap, then they are orthogonal from each other in this representation, since their dot product is zero). 

\begin{figure}
\center
\includegraphics[width=0.5\textwidth]{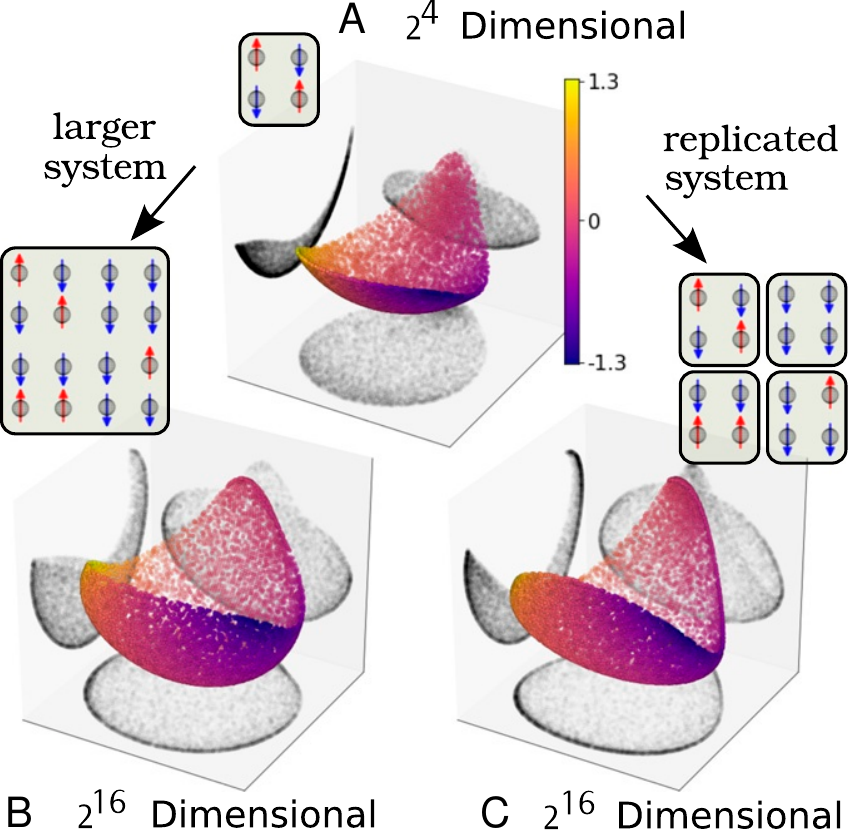}
\caption{
\label{fig:replicas}
{\bf Hypersphere embeddings.} (a)~An original $2\times 2$ Ising model sampled by sweeping through a range of possible field and coupling strengths and visualized using the Hellinger embedding with the first three extracted components of a classic PCA. (b)~As the system size increases to $4\times 4$, the probability distributions are more distinguishable and thus more orthogonal, illustrating the curse of dimensionality. (c)~This effect is simulated by taking four replicas of the $2\times 2$ system. Figure taken from~\cite{Quinn:2019dc}.
}
\end{figure}

The curse of dimensionality that hinders Hellinger can be seen easily in this framework \jps{(see~\cite{Teoh:2020kp})}. As the dimension of the data
$\mathbf{x}$ increases, 
the distributions become more distinguishable from each other.
They therefore become increasingly orthogonal, and all distances between non-identical distributions converge to a constant value as they become maximally apart on the hypersphere. As an illustration of this, consider the finite, two-dimensional Ising model in a field with nearest-neighbor coupling (Fig.~\ref{fig:replicas}). For small system sizes, two similar field and coupling strengths yield similar spin configurations and can be difficult to distinguish from one another. However, as the system size increases (and therefore the dimension of the observed data also increases) the increase in information allows one to more easily evaluate the underlying parameters. The probability distributions representing different parameters are more distinguishable from each other, and are therefore further apart in the hypersphere embedding. In the limit of infinite data, we can completely disentangle all possible parameter combinations: in other words, all distributions become orthogonal and equidistant from one another, forming a hypertetrahedron in probability space. The geometry of a regular hypertetrahedron
with $N$ vertices clearly cannot be captured with fewer than a
$N-1$-dimensional embedding in a Euclidean space. But this does not remain
true if we allow for both space and time-like coordinates\dots

\subsection{Intensive embeddings and replica theory}
\label{subsec:InPCA}

If instead, we look at the information density rather than the total amount of information in a distribution, we can avoid this orthogonality catastrophe by reducing rather than increasing the dimensionality. To accomplish this, we can make use of replica theory. We can simulate the curse of dimensionality by considering replicas of the original system, \textit{i.e.} multiple, independent samples of the original distributions. As an illustration, consider the two-dimensional Ising model with field and nearest neighbor interactions. A $2\times 2$ spin system, visualized in this hypersphere embedding, has $2^4$ dimensions, one for each possible spin configuration. By considering a range of possible field and coupling strengths, we sweep out a region of the model manifold. Figure~\ref{fig:replicas}(a) shows the first three components of a classic PCA of this embedding, with the curvature of the hypersphere very evident. As the system size increases to $4\times 4$ spins, the dimensionality increases and the orthogonality problem becomes more evident. This effect is simulated by looking at 4 replicas of the $2\times 2$ system, shown in Fig.~\ref{fig:replicas}(c).

Rather than increasing the system size with many replicas we decrease it by considering the limit of zero replicas. In this way, we obtain an \textit{intensive} embedding. The distance between points becomes characterized not by the Hellinger distance but the Bhattacharyya%
  \footnote{\jps{The factor of eight is added to match the local distances
  to that of the Fisher Information metric.}}
divergence~\cite{bhattacharyya1946measure}
\jps{
\begin{equation}
\label{eq:Bhat-defn}
d^2_\mathrm{Bhat}[\theta,\tilde\theta] = -8\log\left(\sum_x \sqrt{p(x|\theta) p(x,\tilde\theta)}\right).
\end{equation}
}
By working through the derivation of PCA using the Bhattacharyya divergence, which does not respect the triangle inequality, we obtain an embedding which is pseudo-Riemannian, called InPCA~\cite{Quinn:2019dc}.

\begin{figure}
\center
\includegraphics[width=0.5\textwidth]{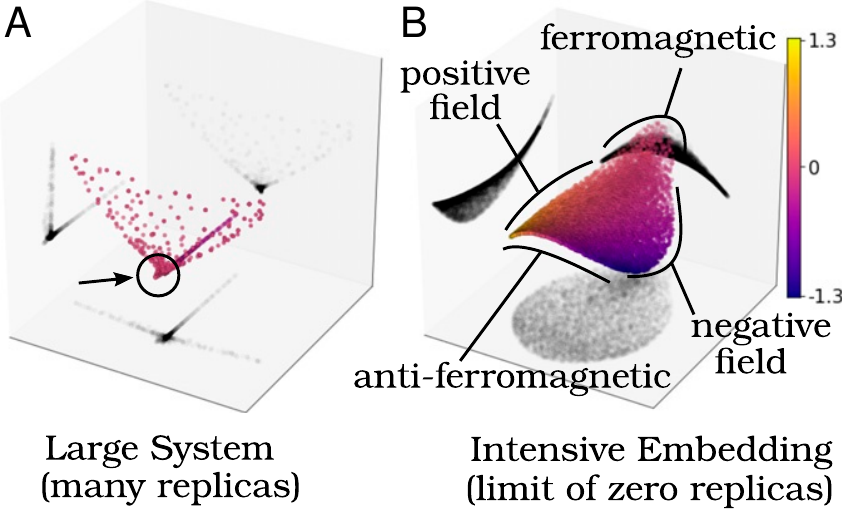}
\caption{
\label{fig:intensive}
{\bf Intensive embedding contrasted with high-dimensional.} (a)~A very high-dimensional visualization of the original $2\times 2$ Ising system by considering multiple replicas, simulating the curse of dimensionality where all points become effectively orthogonal obscuring useful features. (b)~By considering the limit of zero replicas, we extract an intensive embedding (InPCA) which reveals relevant information in the model manifold (such as the ferromagnetic and anti-ferromagnetic regimes). An important difference between (a) and (b) is that the intensive embedding makes use of a pseudo-Riemannian space, with the z-component having a negative-squared distance. Figure taken from~\cite{Quinn:2019dc}.
}
\end{figure}

\begin{figure}
\center
\includegraphics[width=0.7\textwidth]{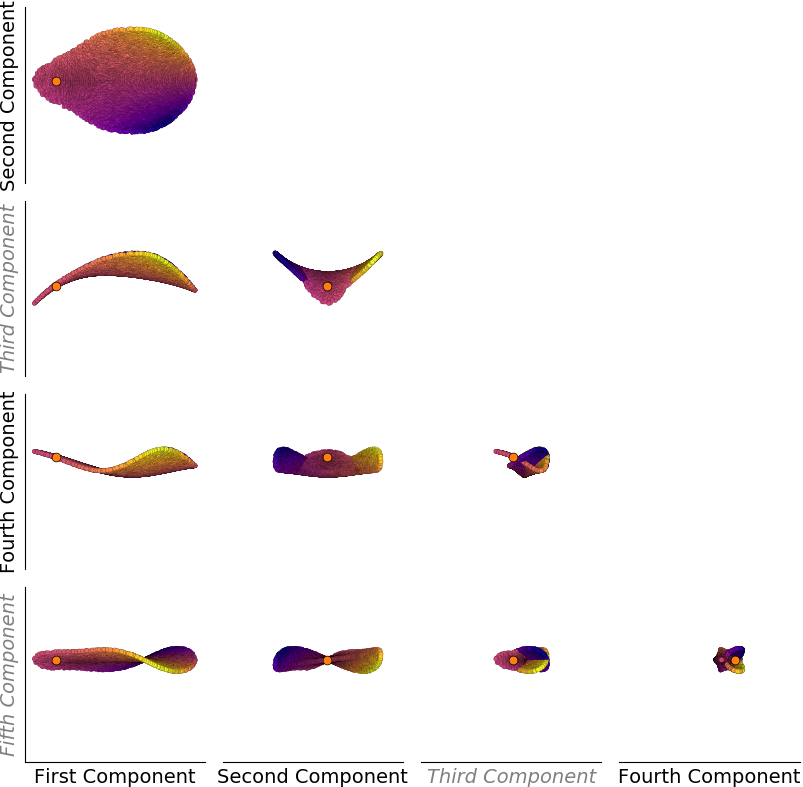}
\caption{
\label{fig:isingTriangle}
{\bf InPCA projections of the $2\times 2$ Ising Model.} The hyperribbon structure of the manifold is apparent, as the successive widths decrease geometrically (widths shown in Fig.~\ref{fig:sloppyLengths}). Color code represents field strength, matching Figs.~\ref{fig:replicas} and~\ref{fig:intensive}. Orange dot represents critical point. Importantly, some components reflect a positive squared-distance (black labels) while others are a negative-squared distance apart (grey italics). Figure taken from~\cite{QuinnPhD}.
}
\end{figure}

Empirical results from probabilistic models show that they share the hyperribbon structures we find in least squares models. In Fig.~\ref{fig:sloppyLengths}, 
for example, we see geometrically decreasing widths for the two-dimensional
Ising model, variable-width Gaussians, CMB skymaps, and neural networks~\cite{Quinn:2019dc,Teoh:2020kp}.  Fig.~\ref{fig:isingTriangle} shows the model manifold for a $2\times2$ Ising model, exhibiting projections along the first five
InPCA components, clearly showing this hyperribbon decrease of widths.

\begin{figure}
\center
\includegraphics[width=0.8\textwidth]{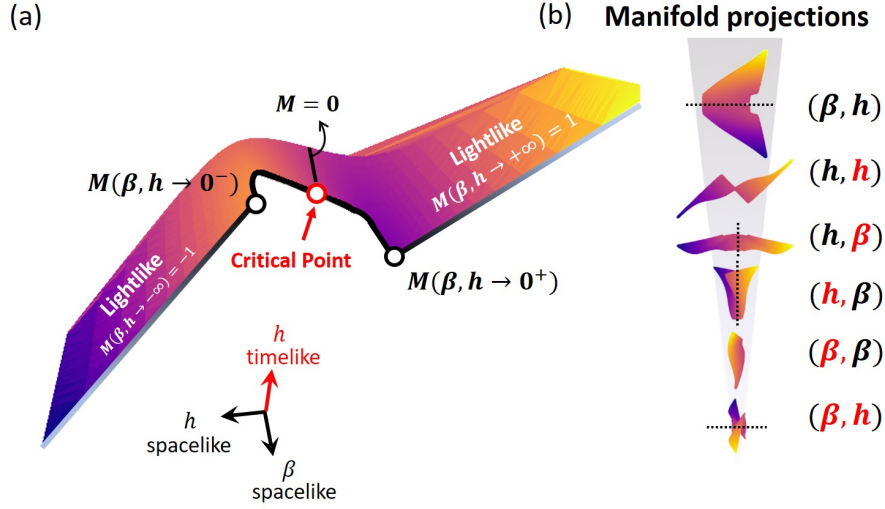}
\caption{
\label{fig:isKLIsing}
{\bf isKL embedding for the 2D Ising model.} 
The coordinates 
$h = H/2 T \pm M/2$ and $\beta = 1/2T \pm E$. $H/T$ and $1/T$ are
the natural parameters where $T$ is the temperature and $H$ the external
field; $M$ and $E$ are the magnetization (total spin) and coupling energy
(sum of the products of neighboring spins). Black coordinates are space-like,
red are time-like (representing negative squared distance).
(a)~ A three-dimensional image of the 4D manifold in Minkowski space. 
The two ``front edges'' represent the fully up-spin and down-spin
configurations, which are unintuitively spread out into a light-like line
(zero net distance) as the external field varies. On the other hand,
it adds intuition to the jump in the magnetization as the field crosses zero;
the $\pm M$ branches are nicely separated by a light-like displacement,
faithfully representing the scaling behavior at the critical point.
(b)~2D manifold projections along the six pairs of coordinates. 
Figure taken from \cite{Teoh:2020kp}.
}
\end{figure}

\subsection{Symmetrized Kullback-Leibler methods}
\label{subsec:isKL}

The Hellinger distance and Bhattacharyya divergence belong to a larger class of \jps{divergences~\cite{csiszar2004information,renyi1961measures}} whose metric is given by the Fisher Information. The most commonly used distance measure for probability distributions, $p(x|\theta) \equiv p$ and $p(x|\tilde{\theta}) \equiv  \tilde p$, is the Kullback-Leibler divergence~\cite{kullback1951information} or relative entropy
\begin{equation}
\label{eq:KL-defn}
D_\mathrm{KL}[\theta, \tilde{\theta}]
    = \sum_x p(x|\theta)\log\left(\frac{p(x|\theta)}{p(x|\tilde{\theta})}\right).
\end{equation}
Because it is asymmetric, often researchers 
\jps{symmetrize~\cite{jeffreys1939theory,kullback1951information}}
the Kullback-Leibler divergence
by considering the sum of the two permutations, which we write as the square of a distance $d_\mathrm{sKL}$:
\begin{align}
\label{eq:symkl}
d^2_\mathrm{sKL} = D_\mathrm{KL}\big[\theta,\tilde\theta\big] + D_\mathrm{KL}\big[\tilde{p} \big\Vert p\big]  
    = \sum_x \Big(p(x|\theta) - p(x|\tilde{\theta})\Big)\log\left(\frac{p(x|\theta)}{p(x|\tilde{\theta})}\right)
\end{align}
By combining the method of projecting data along orthogonal direction from MDS~\cite{Quinn:2019dc} (Section~\ref{subsec:MDS})
with the symmetric KL divergence, we can derive the isKL embedding~\cite{Teoh:2020kp}. An astonishing feature of this method is that it can map exponential families into a {\em finite dimensional} embedding space, with dimension twice the number of model parameters. 

Exponential \jps{families~\cite{nielsen2009statistical}} have probability distributions that can be expressed in the
form familiar to physicists, as a partition function described by a Boltzmann-like distribution:
\begin{align}
\label{eq:exp}
p(x|\theta) = h(x)\exp\Big(\sum_i \eta_i(\theta) \phi_i(x) -A(\theta)\Big),
\end{align}
where $\eta_i(\theta)$ is a function of model parameters and is known as the $i$th natural parameter, $\phi(x)$ is a function only of $x$ and known as the $i$th sufficient statistic, $h(x)$ is a normalizing function, and $A(\theta)$ is
the log partition function.
By combining Eqs.~\ref{eq:symkl} and~\ref{eq:exp}, we see that:
\begin{align}
d^2_\mathrm{sKL} = \sum_i \left( \eta_i(\theta) - \eta_i(\tilde{\theta})\right)\big( \left<\phi_i\right>_\theta - \left<\phi_i\right>_{\tilde{\theta}} \big),
\end{align}
where $\left<\cdot\right>_\theta$ reflects taking an average over $x$ using $p(x|\theta)$. Rearranging the terms in the above equation, we can see that:
\begin{align}
d^2_\mathrm{sKL} &= \frac{1}{4}\sum_i\left( [\eta_i(\theta) + \left<\phi_i\right>_\theta] - [\eta_i(\tilde{\theta}) + \left<\phi_i\right>_{\tilde{\theta}}]\right)^2 - \left( [\eta_i(\theta) - \left<\phi_i\right>_\theta] - [\eta_i(\tilde{\theta}) - \left<\phi_i\right>_{\tilde{\theta}}]\right)^2 
\nonumber
\\
&= [\mathcal{S}_i(\theta) - \mathcal{S}_i(\tilde{\theta})]^2 - [\mathcal{T}_i(\theta) - \mathcal{T}_i (\tilde{\theta})]^2,
\end{align}
where $\mathcal{S}_i$ and $\mathcal{T}_i$ represent space-like and time-like components, and can be directly calculated from the model:
\begin{equation}
\label{eq:super}
\begin{aligned}
\mathcal{S}_i(\theta) &= \frac{1}{2}\big(\eta_i(\theta) + \left<\phi_i\right>_\theta\big), \\
\mathcal{T}_i(\theta) &= \frac{1}{2}\big(\eta_i(\theta) - \left<\phi_i\right>_\theta\big).
\end{aligned}
\end{equation}
There are a {\em finite} number of components, determined by the number of parameters, regardless of the dimension of $x$. This provides a systematic, computationally efficient way of visualizing model manifolds, as opposed to having to calculate the distance between all points and computing the projections ~\cite{Teoh:2020kp}. 
For example, calculating the entire probability distribution
for the Ising model (needed for the Hellinger and Bhattacharyya embeddings) 
becomes infeasible for all but tiny system sizes 
(Figs.~\ref{fig:replicas}--\ref{fig:isingTriangle}), but traditional
Monte-Carlo methods can be used to calculate the magnetization and 
interaction energy needed for the isKL embedding of the Ising model
for any reasonable system size (Fig.~\ref{fig:isKLIsing}).

Moreover, isKL provides a valuable check for data without a known model. By calculating the $d^2_\mathrm{sKL}$ between data, and visualizing both the real and imaginary components extracted using Multidimensional Scaling, we can check to see if the resulting embedding is finite or infinite. If finite, then one is led to believe that the data results from a process described by an exponential family (basically a statistical sampling described by a free energy).

\subsection{Intensive embedding procedure: InPCA, isKL, and MDS}
\label{subsec:MDS}

Here we discuss the methods used to generate the visualizations produced in this section. Classical Principal Component Analysis (PCA) and Multi-dimensional Scaling (MDS~\cite{torgerson1952multidimensional,Quinn:2019dc}) take data sets and infer features which are linearly uncorrelated. PCA allows researchers to examine data in a new basis by centering it and then extracting components of maximal variance, or diagonalize its ``moment of inertia", then extract projections in a hierarchical manner which minimizes the error in distances between points when restricted to a smaller dimensional subspace. Because PCA optimizes distances, one can extend it to take as inputs not the raw data but rather the distances themselves, and in this way obtain MDS. This approach can be generalized to intensive embeddings (InPCA and isKL) in the following way. First, given a set of $N$ probability distributions $\bx_i$, calculate the squared distance, \textit{i.e.} the Bhattacharyya divergence or Symmetrized Kullback-Leibler divergence between them, $X_{i,j} = D(\bx_i \mid \mid \bx_j)$ where $D$ represents the choice of divergence. Next, generate the intensive cross-covariance
matrix~\cite{Quinn:2019dc}, $W = - PXP/2$, where $P_{ij} = \delta_{ij} - 1/N$ serves to mean-shift the data, so that it is centered on its center of mass.

Next, compute the eigenvalue decomposition for $W$, $W=U\Sigma U^T$, and note that there may be both positive and negative eigenvalues. It is important that they be ranked by magnitude, \textit{i.e. $|\Sigma_{00}| \geq |\Sigma_{11}| \geq \dots$}. The coordinate projections for the intensive embedding are determined by;
\begin{align}
T = U\sqrt{\Sigma}.
\end{align}
One then plots the projections using the columns of $T$. Coordinates corresponding to positive eigenvalues represent space-like components, in that they have a positive-squared contribution to the distance between points. Conversely, coordinates corresponding to \textit{negative} eigenvalues represent time-like components, in that they have negative-squared contributions to the distance between points. The resulting figures are interpreted much like a spacetime diagram, where points with zero distance are light-like separated. Note that, because all points are a positive-squared distance apart (since their divergence is non-negative) their overall distance will be a space-like separation, although they may be visualized using both space-like and time-like components.

\section{Future directions}

In this review, we discussed how information geometry explains sloppiness
in multiparameter models: predictions are independent of all but a few
parameter combinations, which is unfortunate if you want to deduce the
parameters from collective behavior, but fortunate if you want to test the
predictions of the model before the parameters are fully measured. Extensive
observations~\cite{Gutenkunst:2007gl} and rigorous results~\cite{Quinn:2018tw} explain this by noting that the predictions
form a hyperribbon structure. We've described the MBAM method~\cite{Transtrum:2014hr}
for deriving simpler, emergent models compatible with the data using the 
boundaries of the model manifold along thin axes. We've introduced new Bayesian priors~\cite{Mattingly:2017uao}, that implement Occam's razor by placing
$\delta$-function preferences 
for these simpler boundary models. We've connected the hyperribbon structure
of the model manifold in emergent statistical mechanics models to 
renormalization-group flows~\cite{Raju:2017ty}. And we've introduced visualization methods
that bypass the curse of dimensionality for high-dimensional data sets by
generating isometric hyperribbon embeddings in Minkowski-like spaces~\cite{Quinn:2019dc,Teoh:2020kp}.

What is there left to do? In previous reviews and papers, we have explored and speculated about applications of information geometry to systems biology~\cite{Gutenkunst:2007gl}, power systems~\cite{transtrum2017measurement}, robustness and neutral spaces in biology~\cite{Daniels2008}, controlling and optimizing complex instruments such as particle accelerators~\cite{BerganBCLRS19}, and explaining why science works~\cite{Transtrum:2015hm}. Here we focus on possible future developments related to the new tools and methods discussed here -- understanding why hyperribbons and emergence arise, using their boundaries as simpler models, relations to emergence in physics, new Bayesian priors, and visualization methods that bypass the curse of dimensionality.


\subsection{Visualization, experimental design, and testing model validity}

The sloppiness of multiparameter models suggests that extracting parameters from data will be challenging or impossible. Model reduction methods such as MBAM show that one can describe data in complex systems with emergent models for which many parameters and model components are simply ignored (set to zero, infinity, or replaced by other parameters), {\em even when those components of the model are microscopically known to be important and relevant.} While we may find the emergent models more comprehensible, aren't they wrong? The rates in a system biology model do not go to infinity, the enzymes are not in the limit of short supply -- the simpler model works by shifting other parameters to compensate for these simplifying assumptions.

The progress of science often exhibits this in reverse -- one develops a simple theory which captures the essential features of a problem, and then adds complexity to address new experimental facts.  Can we use information geometry to guide the design of new experiments to pin down models in an optimal way? Optimal experimental design~\cite{CaseyBFGWMBCS07} is a well-developed branch of statistics, but usually is done in a linearized approximation (but see~\cite{vanlier2012bayesian}) -- can we use hyperribbon embeddings and differential geometry to turn experimental design into a more global tool? As a particular example, the InPCA methods were developed in order to visualize the space of possible Universes as reflected by current measurements of the microwave background radiation. The next generation of instruments are designed to measure and test new physical phenomena -- gravitational waves from the epoch of inflation, evidence of light relic particles in the early universe, and signatures of dark matter in the cosmic microwave background radiation. Would it be useful to visualize the predictions of extensions of the $\Lambda$CDM model that include these phenomena, by mapping out those portions of the model manifold that are consistent with current experimental data? Can we use such visualizations to help guide the choice of the next generation of instruments?

Our information geometry gives us a theoretical justification for trusting the predictions of nonlinear models with many undetermined parameters. Climate science, for example, incorporates not only well-controlled theories of heat, hydrodynamics, and radiation, but much more speculative models of clouds, glaciers, and forest ecologies (not to mention human behavior). It is reassuring to hear that equally complex models in systems biology with even more undetermined parameters are quite useful in guiding drug design (see below). But just because complex models fitted to data can make correct predictions does not mean they must do so. Statistics provides many tests of model validity and many methods to compare the predictive power of rival models, for cases where the parameters are well determined. Can we use information geometry methods to create similar tests and comparison tools for multiparameter models where the uncertainties in model parameters are typically not only outside the linear regime, but often unbounded?

\subsection{Understanding emergent low-dimensional behaviors}
Emergence in physics relies on small parameters. Continuum limits work (elasticity, fluids, superfluids\dots)
when frequencies and wavelengths are low; the renormalization group works
(emergent scale invariance, avalanches, crackling noise, \dots)
when in addition there is a proximity to a continuous transition 
between states. Sections~\ref{sec:ParameterCompression} and~\ref{subsec:isKL}
discussed how information geometry applied to these models form hyperribbons.
On the other hand, we used interpolation theory to explain the
low-dimensional emergent behavior and reduced models that arise in nonlinear
least-squares models fit to data (Section~\ref{sec:HyperribbonBounds}).
Emergence in physics and emergence in least-squares models are both 
characterized by sloppiness (only a few combinations of microscopic
parameters dominate the behavior), and both form
hyperribbons, but our explanations for the two seem completely different.
For more general probabilistic models, intensive embeddings (Section~\ref{sec:MinkowskiSpace}) usually lead to hyperribbons and emergent theories,
but we do not yet have an explanation for this in general, except for
the special (but very useful) case of exponential families, where the model 
manifold not only is thinner and thinner along successive cross sections, 
but becomes zero thickness after a finite number of embedding dimensions.

Can we find a common theme between these disparate results -- a unified
theory for the emergence of simplicity from complexity? One possible clue
is that sloppiness is a property both of the scientific model and the 
predictions of interest. The Ising model is not sloppy, and has no beautiful
emergent theory, unless one only cares about long length and
time scales. And a systems biology model is not sloppy if one designs
experiments to measure one parameter at a time.
Multiparameter least squares models have emergent simplicity  when one
focuses on collective behavior of the system as a whole, and statistical
mechanics has emergent simplicity when one focuses on collective behavior
of large numbers of particles. Perhaps it is caring about the answers to the
right questions that makes science possible?

\subsection{Intensive embeddings for big data, from biology to deep neural networks} 

In our work visualizing the model manifold for probabilistic models (Section~\ref{sec:MinkowskiSpace}), we
grappled with the {\em curse of dimensionality}, commonplace in big data
applications. We identified the problem of having too much information:
data points in too high a dimensional space naturally becoming orthogonal.
We cured the problem first using replica theory and then
using an analytic method for models that form `exponential families'.
Our cures work because of two features. (1)~We found a way of embedding
our model in a Minkowski-like space, using MDS (multidimensional scaling
(MDS~\cite{Quinn:2019dc}, Section~\ref{subsec:MDS}), that violates the
triangle inequality. (2)~We use measures of squared distance between
data points (Bhattacharyya and symmetrized Kullback-Leibler) that
diverge logarithmically with the overlap between the probability
distributions. (This ensures that doubling the amount of data simply
rescales all distances by a factor of two). 

Can we apply these basic ideas to other big data problems, that are not framed as probabilistic models?
We know of at least one machine learning method, PHATE~\cite{Moon19} (due to Kevin Moon) that has applied these two principles to visualize biological
data (in particular single-cell RNA sequencing data sets). 
Here let us consider two possible applications to machine learning, as
represented by the millions-of-parameter deep learning neural network models.

Deep neural networks are `big' in the number of neurons per layer (hundreds to thousands), the number of neural connections between layers (millions of weight parameters), the number of dimensions in the initial data used to train the network (say, one per pixel in the images being recognized), and the number of data points used to train the network (millions of images). They have been remarkably successful at extracting meaningful information from these large data sets, suggesting that there is some hidden, low-dimensional structures~\cite{GoldtMKZ20} in the data, that is disentangled and flattened via the nonlinear transformations through succeeding neural layers.

With Pratik Chaudhari, some of us have been exploring~\cite{PratikChaudhariWIP}
how one might use
intensive embeddings to visualize both the training of these neural networks (as in~\cite{Quinn:2019dc}) and potentially explore low-dimensional structures discovered after the training process. Consider a network that categorizes,
taking $N$ images $\bx$ and returns a preference probability $\brho(\bx)$ whose
largest value is the predicted category. The vector $\bR =(1/N) \{\brho(x_1), \brho(x_2), \dots\}$ can be viewed as a probability distribution over the whole
set, and one could use InPCA or isKL to visualize how the network trains with time, averaging over the random initial states and training steps. Visualization of the resulting structures for training and test data
could lend geometrical insight into how and why different training protocols differ in their generalization errors. Alternatively, one could examine the 
neural activations $h(\bx)$ of the different layers of a single neural network
-- each encoding distilled information about the initial data $\bx$ gleaned from
the previous layer that is most relevant to the task at hand. Using PCA on the
vectors $h(x)$ is likely doomed for the standard reasons, but conceivably
using MDS with an intensive metric could allow one to see the emergence of
a low dimensional (perhaps hyperribbon-like) representation of the data as
one passes through successive trained layers.

\subsection{Possible applications of model reduction}

The manifold boundary approximation method is a principled way to select effective theories for sloppy systems.
MBAM is objective, data-driven, and does not require expert intuition.
Through sequences of simplifying approximations, it gives identifiable models that retains essential elements, such as dominant feedback loops, while remaining expressed in terms of microscopic physics.
We anticipate that simplified models will be increasingly important by enabling clearer reasoning about big data and complex processes.

Recent result from systems biology modeling the Wnt pathway illustrate how this may be done.
Wnt signaling is the primary mechanism by which short-range extra-cellular signals are relayed to the nucleus, and it plays a major role in normal embryonic development as well as cancer tumerogenesis.
While the real system is notoriously complex, it regularly exhibits simple, comprehensible behavior.
For example, the steady state conditions of the pathway may encode for the level of extra-cellular Wnt~\cite{goentoro2009evidence} while at other times, a negative feedback drives sustained oscillations that regulate somitogenesis~\cite{jensen2010wnt}.
In each case, the behavior is explicable in terms of simple models with only a few basic mechanisms.
However, by reasoning about the relationships between these reduced models within the larger space of candidate approximations, more is possible.
The \emph{supremal model}, i.e., simplest model reducible to each effective behavior includes the relevant mechanisms for \emph{transitioning} between states~\cite{petrie2021supremum}.
This model makes predictions that are qualitatively different from the data to which it was trained and is a starting point for controller design.

Can simple, effective models facilitate the transition of big data into practical knowledge in spite of being incomplete?
Quantitative Systems Pharmacology (QSP) uses complex models to design and engineer drugs~\cite{knight2016promises}, but reasoning about causality within complex models is difficult.
In our conversations with the FDA and pharmaceutical companies, challenges in justifying new drugs and human trials are closely related to model complexity~\cite{transtrumfda2020}.
If it is unreasonable to identify all of the parameters in realistic biological models, how can drug developers meet standards of government approval agencies~\cite{peterson2015fda}?
Simple models that are rigorously derived from first principles can be more easily and cheaply validated.
Furthermore, simple models that contain only the relevant mechanism are more readily communicated and trusted in ways that complicated simulations are not.

Can similar reasoning be applied beyond biology?
Empirical potentials in molecular modeling are designed to mimic quantum effects within a classical framework.
But which combination of quantum effects are responsible for specific macroscopic material properties?
The combinatorial large space of stoichiometry and material properties make the problem particularly challenging.
Large databases of quantum calculations~\cite{curtarolo2012aflow, saal2013materials} archive data while projects such as OpenKIM~\cite{tadmor2011potential} satisfy the dual problem of documenting and archiving useful models.
But leveraging these data and models to design new materials remains an open problem.
Can simple models that are successful in a limited context help clarify what must be accurately modeled to guide material by design?

We have speculated about the relationship between sloppiness and large machine learning models.
Although model size seems to play an important role in learning capacity and training rate, large models cannot be deployed on small devices without a reduction step.
Furthermore, large black-box models also do not lead to explicit, declarative knowledge in the same way as minimal, parsimonious representation~\cite{holzinger2018machine}.
The nascent fields of explainable AI~\cite{hagras2018toward} and physics-guided machine learning~\cite{pawar2021physics, willard2020integrating, rai2020driven} may also benefit from simplification in a context-adaptive way.

We have also mentioned climate as a field whose predictions are the result of multi-physics simulations.
Uncertainties in specific components (e.g., models of clouds or forest ecology), as well as their role in the larger simulation, make it difficult to reason about causality in such simulations.
Simplified models that automatically identify the essential feedback loops and effective control knobs may guide how modelers understand how these uncertain components affect conclusions.
They may also aid science communication and education for policy makers and the general public.

The situation in climate is reminiscent of that in power systems that similarly use multi-physics to model large engineered systems.
When complex models fail to match data (even presumably identified models), for example in well-instrumented blackouts~\cite{andersson2005causes}, it is bewildering to identify the source of the problem.
Simplified models make it easier to identify and correct errors in the model.
Our model reduction methods have made inroads towards this end in power systems~\cite{transtrum2017measurement}; can they similarly find use in these other areas?

Power systems application also face the challenge that they operate at the intersection of industry and government where issues of individual privacy, proprietary information, and public security are relevant.
Model reduction is a means to limit and regulate how information is shared or redacted among vested parties.
Models that contain sensitive infrastructure information can be systematically and objectively coarse-grained to remove classified pieces while simultaneously meeting the demands of public disclosure and oversight.

The computational cost of solving the geodesic equation limit the scope of current algorithms to models with at most a few hundred parameters, so alternative strategies are needed.
It may be possible to guide optimization algorithms toward the boundary; merging the fitting and reduction steps.
Alternative approaches may focus on targeted subsets of parameters.
The topological relationship among potential models remains largely unexplored, but such a theory could better inform model selection for target applications.

\subsection{Conclusions}

Efficiently reasoning with big models and big data will require new organizational paradigms~\cite{laughlin2000cover, crutchfield2014dreams}.
The physical theories of the twentieth century, interconnected through a web of classical limits as in Figure~\ref{fig:newtoneinsteinschrodinger}, suggest a way forward.
Effective theories organize our understanding and focus our attention to the relevant physical principles within specific regimes of validity.
The hyperribbons presented here reflect a similar information hierarchy, but sloppiness adds a new wrinkle to the traditional narrative by justifying simple theories without resorting to classical limits, coarse-graining, or renormalization.
Visualization tools and model reduction methods extract simple explanations in a context-specific way, and our theories of information topology organize them into a network of interrelated models, each applicable at an appropriate scale and for specific applications.
While science has traditionally progressed by discovering effective theories before the underlying mechanism, the future of science may well operate in the opposite direction, cataloging and organizing effective theories for the universe's myriad complexities.
Sloppy models abetted with the formalism of information geometry explain why and how this can be done.





\section*{Acknowledgments}

We would like to thank David Schwab and Pratik Chaudhari for helpful conversations.
KNQ was supported in part by the US National Science Foundation, through the Center for the Physics of Biological Function (PHY-1734030) and from NSF DMR-1719490. Parts of this work were performed at the Aspen Center for Physics, supported by NSF grant PHY-1607611. The participation of MCA at Aspen was supported by the Simons Foundation. MKT was supported by NSF EPCN-1710727, NSF CMMT-1753357, and NSF CMMT-1834332. JPS was supported by NSF DMR-1719490. BBM and MCA were supported by a Simons Investigator award, and BBM was supported by NIH R35 GM138341

\section*{References}
\bibliographystyle{my-JHEP-4.bst}
\bibliography{MultiparameterModels}

\end{document}